\documentclass[11pt]{article}

\fussy 
\usepackage{amssymb}
\usepackage{epsf}
\usepackage{graphicx}

\begin{document}
\ifx\href\undefined\else\hypersetup{linktocpage=true}\fi 

\title{Flatland Electrons in High Magnetic Fields}
\author{M. Shayegan \\
\textit{Department of Electrical Engineering}\\ \textit{Princeton
University, Princeton, New Jersey USA}}

\maketitle \tableofcontents

\newpage

\section{Introduction}

Electrons in a ``flatland'' are amazing! A simple low-temperature
measurement of the resistance of a two-dimensional electron system
(2DES) as a function of perpendicular magnetic field $\left(
B\right) $ reveals why (Fig. 1). In this figure the resistivities
along $\left( \rho _{xx}\right) $ and perpendicular $\left( \rho
_{xy}\right) $ to the direction of current
are shown, and the vertical markings denote the Landau-level filling factor (%
$\nu $). Look how the behavior of $\rho _{xx}$ with temperature
($T$), shown schematically in the inset, changes as a function of
the magnetic field. At certain fields, marked \textbf{A}, $\rho
_{xx}$ drops exponentially with decreasing temperature and
approaches zero as $T\rightarrow 0$. This is the quantum Hall effect
(QHE) and, as you can see in the other trace of Fig. 1, the Hall
resistance $\left( \rho _{xy}\right) $ becomes quantized near these
fields. The QHE is best described as an incompressible quantum
liquid which can possess a high degree of short-range electron
correlation (e.g., when the QHE occurs at a fractional $\nu$). Next,
look at the $T$-dependence of $\rho _{xx}$ at the fields marked
\textbf{B}
(near 13 and 14 T for this sample). Here $\rho _{xx}$ exponentially \emph{%
increases} with decreasing $T$, signaling an insulating behavior.
The nature of this insulating state is not entirely clear, but it is
generally believed that it is a pinned Wigner solid, a ``crystal''
of electrons with long-range positional order (see Fig. 2). Now look
at what happens at the magnetic field marked \textbf{C}. At this
field, $\rho _{xx}$ shows a nearly temperature-independent behavior,
reminiscent of a metal. It turns out that at this particular field
there are two flux quanta per each electron. The electron magically
combines with the two flux quanta and forms the celebrated
``composite Fermion," a quasiparticle which now moves around in the
2D plane as if no external magnetic field was applied!
\begin{figure}\centering
\includegraphics[scale=0.3]{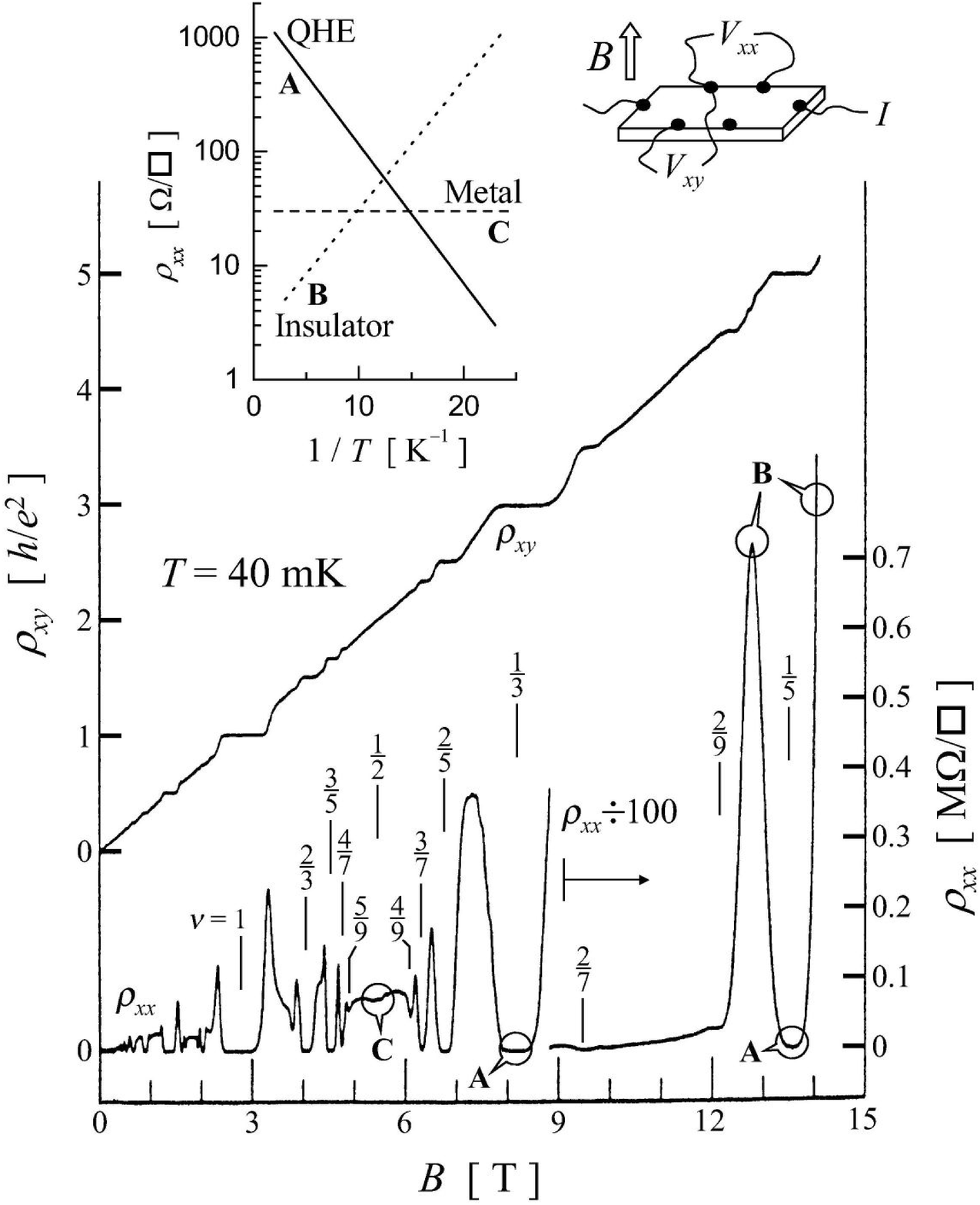}
\caption {Low-temperature magnetotransport coefficients of a
high-quality (low-disorder) 2D electron system in a modulation-doped
GaAs/AlGaAs heterostructure with a 2D density of $6.6\times 10^{10}$
cm$^{-2}$. The longitudinal $(\rho _{xx})$ and Hall $(\rho _{xy})$
resistivities at a temperature of 40 mK are shown in the main
figure. The Landau-level filling factors $(\nu )$ are indicated by
vertical markings. The right upper inset shows the typical
measurement geometry while the left inset schematically illustrates
the widely different temperature dependences of $\rho _{xx}$ at
different magnetic fields (filling factors), marked by \textbf{A},
\textbf{B}, and \textbf{C} in the main figure. (After Sajoto
\emph{et al.} \cite{Sajoto1993}.)}
\end{figure}
\begin{figure}\centering
\includegraphics[scale=0.5]{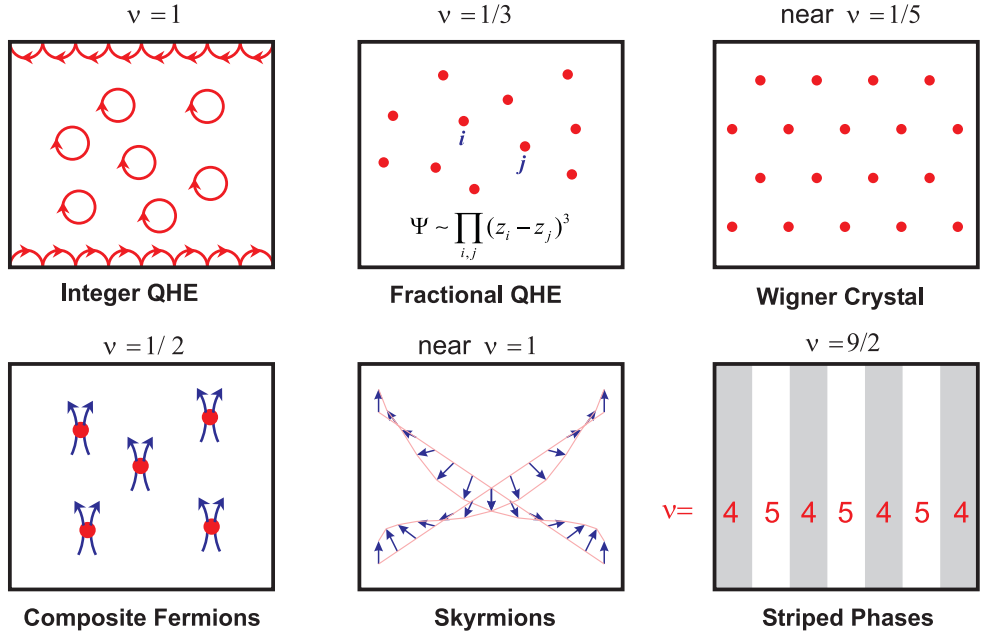}
\caption{Some of the different states of a low-disorder 2D
electron system in a strong perpendicular magnetic field.  The
only parameter that is changing is the Landau level filling factor
$(\nu )$ which is inversely proportional to the magnetic field.
Except for the integer QHE, all the other states are stabilized by
the electron-electron interaction.}
\end{figure}
\begin{figure}\centering
\includegraphics[scale=0.25]{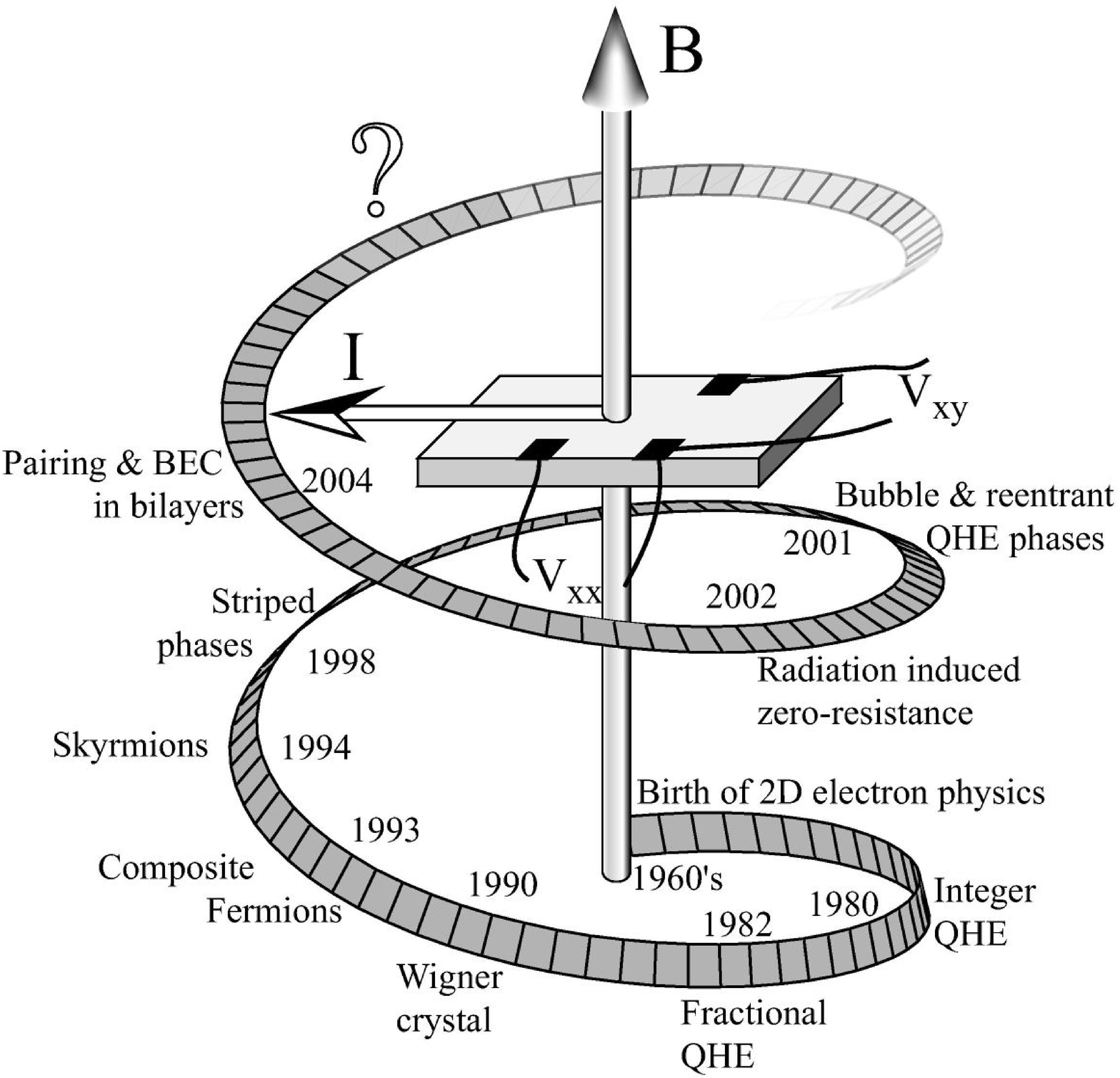}
\caption{Some of the noteworthy discoveries in the field of flatland
electrons in a perpendicular magnetic field.}
\end{figure}

So in one sweep, just changing the magnetic field, the 2DES shows a
variety of ground states ranging from insulating to metallic to a
``superconducting-like" phase.  But wait, that's not all!  During
the past decade, yet more new phases and phenomena have been
discovered (see Figs. 2 and 3). For example, near certain magnetic
fields, the spins of electrons have a remarkable texture, as the
so-called ``Skyrmions" are present.  Yet at other fields, the ground
state is a ``striped phase" where the electron density is modulated
in one direction and the electron transport in the plane becomes
extremely anisotropic.  As it turns out, these ground states are all
stabilized primarily by strong electron-electron correlations. The
presence of so many novel states attests to the extreme richness of
this system, one which has rendered the field of 2D carrier systems
in a high magnetic field among the most active and exciting in solid
state physics. It has already led to two physics Nobel prizes, one
in 1985 to K. von Klitzing for the discovery of the integral QHE
(IQHE) \cite{Klitzing1980}, and another in 1998 to R.B. Laughlin,
H.L. Stormer, and D.C. Tsui for the fractional QHE (FQHE)
\cite{Tsui1982,Laughlin1983}, but surprises don't seem to stop.

The purpose of this article is to provide a glimpse of some of the
exciting experimental results in this field.  My presentation will
approximately follow the history shown in Fig. 3 and will focus on
the following areas:
\begin{enumerate}
\item  a quick summary of some of the sample parameters and
experimental aspects;

\item  some basic and general remarks on the different states of a
2DES in a strong \emph{perpendicular} magnetic field, including the
QHE, Wigner crystal, composite Fermions, Skyrmions, and striped
phases;

\item  \emph{bilayer} electron systems in which the additional
(layer) degree of freedom leads to unique QHE and insulating states
which are stabilized by strong intralayer \emph{and} interlayer
correlations. A highlight is the recent observation of pairing of
carriers in two closely-spaced, interacting layers and the
signatures of the Bose-Einstein condensation of the pairs
(excitons).
\end{enumerate}

I'd like to emphasize that this article is far from properly dealing
with all the important and exciting aspects of the physics of 2D
systems in high magnetic fields. It provides only a limited and
selective sample of transport measurements. Readers interested in
more details are referred to the original papers as well as
extensive review articles and books
\cite{EP2DSconfs,Prange1990,Chakraborty1995,DasSarma1997,Davies1998,AndoFowlerStern1982}.
Also, there will be a minimal treatment of theory here; for more
details and insight, I suggest reading the illuminating article by
D. Yoshioka in this volume and various articles in Refs. 5 to 10.

\section{Samples and Measurements}

\subsection{Two-dimensional electrons at the GaAs/AlGaAs interface}

One of the simplest ways to place electrons in a flatland is to
confine them to the interface between two semiconductors which have
different bandgaps. An example is shown in Fig. 4 (left) where a
2DES is formed at the interface between undoped GaAs and AlGaAs
\cite{Davies1998}. The larger bandgap of AlGaAs leads to its
conduction-band energy $\left( E_{CB}\right) $ being higher than
GaAs. The system is ``modulation-doped'' \cite{Stormer1979} meaning
that the dopant atoms (in this case, Si donors) are placed in AlGaAs
at some distance away from the interface. The electrons from the
donors find it energetically favorable to transfer to the lower
energy conduction-band of GaAs. But as they transfer, an electric
field sets up between the positively-charged (ionized) donors in
AlGaAs and the transferred electrons in GaAs. This electric field
limits the amount of charge transfer. Figure 4 (lower left)
schematically shows $E_{CB}$ as a function of position, at
equilibrium, after the charge transfer has taken place
\cite{Davies1998}. An alternative way to form a 2DES is to confine
the electrons in a GaAs quantum well which is flanked by
modulation-doped AlGaAs barriers. This is shown schematically in
Fig. 4 on the right.
\begin{figure}\centering
\includegraphics[scale=0.7]{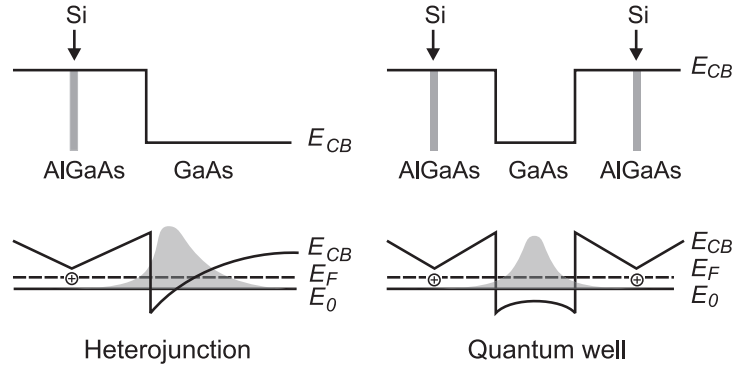}
\caption{Schematic descriptions of modulation-doped GaAs/AlGaAs
samples. Since the conduction-band edge $\left( E_{CB}\right) $ of
GaAs lies lower in energy than that of AlGaAs, electrons transfer
from the doped AlGaAs region to the undoped GaAs to form a quasi-2D
electron system (2DES) at the heterojunction interface between GaAs
and AlGaAs (left), or in a GaAs quantum well (right). In both cases,
the 2DES is separated from the doped AlGaAs region by an undoped
AlGaAs (spacer) layer to minimize electron scattering by the ionized
impurities. The ground state subband energy $E_{o}$ and the Fermi
level $E_{F}$ are shown. Note that the electron wavefunction, $\psi
(\emph{z})$, has a finite extent in the direction perpendicular to
the plane in which the electrons move freely.}
\end{figure}

A key point in the structures of Fig. 4 is that the 2DES is
separated from the ionized dopants. As a result, the scattering of
electrons by the ionized impurity potential is significantly
reduced, meaning that the 2D electrons are essentially ``free'' to
move in the plane. It turns out this is crucial for much of the
phenomena observed in these systems: by reducing the disorder and
the electron-impurity interaction, electrons are allowed to interact
with each other, and the result is a host of new many-body ground
and excited states. Another important message here is that although
we call the system ``two-dimensional,'' the electron wavefunction
$\psi \left(z\right) $ spreads in the $z$ direction by a finite
amount, typically $\sim $ 10 nm. This finite layer-thickness plays
an important role and should be taken into account when comparing
theoretical calculations and experimental results: it distinguishes
between ``ideal'' 2D system assumed in many calculations and the
``real'' quasi-2D, experimental system.

How does one fabricate structures like those in Fig. 4 and what are
the details of a typical sample structure? The best quality
GaAs/AlGaAs samples are presently grown by molecular beam epitaxy
(MBE) \cite{Cho1995}. The MBE system (Fig. 5) is essentially a very
``clean'' high-vacuum evaporation chamber. A GaAs substrate, heated
to about 630${{}^{\circ }}$C, is positioned in front of effusion
cells (ovens) each of which contains one of the required elements
(Ga, Al, As, and Si). The ovens are heated to appropriate
temperatures to produce fluxes of these elements which can impinge
on the GaAs substrate. Each oven also has a shutter which is
controlled, normally via a computer, to produce the desired
structure. Under these circumstances, and with a growth rate of
about one monolayer of GaAs per second (which is roughly 1
$\mu$m/hour), one can grow very high quality, \emph{single-crystal}
structures with nearly any design.
\begin{figure}
\centering
\includegraphics[scale=0.2]{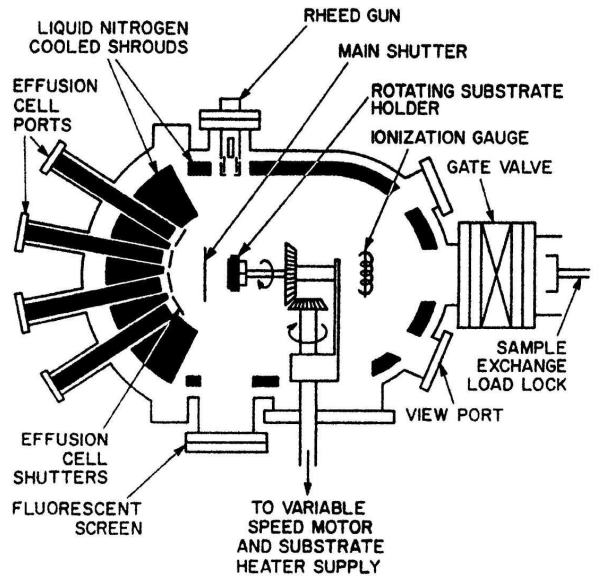}
\caption{Cross-sectional view of a molecular beam epitaxy (MBE)
growth chamber (after Cho \cite{Cho1995}), essentially a very
high-vacuum evaporation chamber with a base pressure of $\sim
10^{-14}$ atmosphere. The chamber is equipped with various vacuum
pumps, such as ion-pumps and cryopumps, and also can have
analytical equipment such as a reflection high-energy electron
diffractometer (RHEED) to monitor \emph{in-situ} the substrate
surface morphology as well as growth rate.}
\end{figure}

What determines the ``quality'' of the 2DES? For the
electron-interaction-dominated phenomena in which we are interested
here, the best sample is typically one with the least amount of
imperfections such as interface irregularities, ionized impurities,
etc. It is this consideration that leads to structures where the
2DES is typically separated from the Si dopants by a very thick
spacer layer of undoped AlGaAs. Details and rationale for other
fabrication procedures such as growth interruptions, the use of a
spacer with graded Al composition, double-$\delta $-doping etc., can
be found in Ref. 13. But a very important factor determining the
quality of the 2DES, one which is not explicitly apparent in the
structures of Fig. 4, is the amount of residual (or unintentional)
impurities that are incorporated \emph{throughout} the structure
during the MBE growth. These impurities are always present because
the vacuum in the MBE chamber is not perfect, and also because the
source materials (Ga, Al, etc.) used in the ovens are not 100\%
pure. It turns out in fact that in structures with a large $(>200$
nm) spacer layer thickness, the most important factor in obtaining
very low-disorder 2DES is the \emph{purity} of the grown material
and not the specific details of the structural parameters. The
vacuum integrity of the MBE growth chamber and the cleanliness and
purity of the source materials and the GaAs substrate are therefore
of paramount importance for the fabrication of state-of-the-art
2DESs.

A measure of the electronic ``quality'' of a 2DES is its
low-temperature mobility, $\mu $. Over the years, the mobility of
modulation-doped GaAs/AlGaAs heterostructures has improved
tremendously and the record stands
at over $10^{7}$ cm$^{2}$/Vs for a 2DES density $\left( n\right) $ of $%
\sim 2\times 10^{11}$ cm$^{-2}$, implying a mean-free-path of tens
of microns \cite{Pfeiffer1989,Recordmobility}. This mobility is more
than $\sim 10^{4}$ times higher than $\mu $ for a uniformly-doped
GaAs layer, demonstrating the striking power of modulation-doping.
As mentioned in the last paragraph, the mobility in such
thick-spacer structures is in fact limited by the concentration of
the non-intentional (residual) impurities. This is evidenced by the
observation \cite{Shayegan1988,Sajoto1990,Kane1993}
that $\mu \sim n^{\gamma }$ with $\gamma \simeq 0.6$%
; this is the dependence expected if the dominant source of
scattering is the \emph{residual} impurities in the close proximity
of the 2DES \cite{Stern1983}. The residual impurity concentration,
deduced from the mobility values for state-of-the-art 2DES with $\mu
\gtrsim 10^{7}$ cm$^{2}$/Vs for $n\gtrsim 5\times 10^{10}$ cm$^{-2}$
is $n_{i}\lesssim 1\times 10^{13}$ cm$^{-3}$, consistent with the
residual GaAs doping expected in very clean MBE systems. An
$n_{i}\sim 10^{13}$ cm$^{-3}$ means that the average distance
between the residual impurities $\sim$ 500 nm is much larger than
the typical inter-electron distance in the 2DES ($\sim 45$ nm for
$n=5\times 10^{10}$ cm$^{-2})$. Clearly in such low-disorder 2D
systems it is reasonable to expect that the physics can be dominated
by electron-electron interaction.

\subsection{Magnetotransport measurement techniques}

A variety of experimental techniques has been used to probe the
electrical, optical, thermal, and other properties of the 2DES in a
high magnetic field. The bulk of the measurements, however, have
been of the magnetotransport properties. Magnetotransport
measurements are also by far the main topic of this paper. I
therefore briefly discuss such measurements here. In typical
\emph{dc} (or low-frequency, $\lesssim 100$ Hz) transport
experiments, the diagonal and Hall resistivities are measured in a
Hall bar or van der Pauw geometry with $\sim  $1 mm distance between
the contacts. Contacts
to the 2DES are made by alloying In or InSn in a reducing atmosphere at $%
\sim 450{^{\circ }}$C for about 10 minutes. High-frequency
measurements often involve more specialized geometries and
contacting schemes. The low-temperature 2D carrier concentration
can be varied by either illuminating the sample with a
light-emitting diode or applying voltage (with respect to the
2DES) to a back- and/or front-gate electrode. Low temperatures are
achieved using a $^{3}He/^{4}He$ dilution refrigerator, while the
magnetic field is provided either by a superconducting solenoid or
a Bitter magnet, or a combination of the two. The low-frequency
magnetotransport measurements are typically performed with a
current excitation of $\lesssim 10^{-9}$ A, corresponding to an
electric field of $\lesssim 10^{-4}$ Vcm$^{-1}$, and using the
lock-in technique.

\section{Ground states of the 2D system in a strong magnetic field}

\subsection{The integral quantum Hall effect (IQHE)}

A large magnetic field applied perpendicular to the plane of a 2DES
acts like a harmonic oscillator potential and leads to the
quantization of the orbital motion. The allowed energies are
quantized and are given by the ``Landau Levels'' (LLs),
$(N+\frac{1}{2})\hbar \omega _{c}$, where $N=0,1,2,...$ and $\hbar
\omega _{c}=\hbar eB/m^{*}$ is the cyclotron energy. For a system
with a finite effective Lande $g$-factor ($g^{*}$), the energy
spectrum is further quantized as each LL is spin-split to two levels
separated by the Zeeman energy $\left| g^{*}\mu _{B}B\right| $ where
$\mu _{B}$ is the Bohr magneton. This evolution of the
density-of-states, $D(E)$, for a 2D system in a magnetic
field is schematically shown in Fig. 6. Note that for 2D electrons in GaAs, $%
m^{*}=0.067m_{o}$ and $g^{*}\simeq -0.44$, so that the cyclotron
energy is about 70 times larger than the (bare) Zeeman energy.
\begin{figure}\centering
\includegraphics[bb=0 45 500 703, scale=0.5]{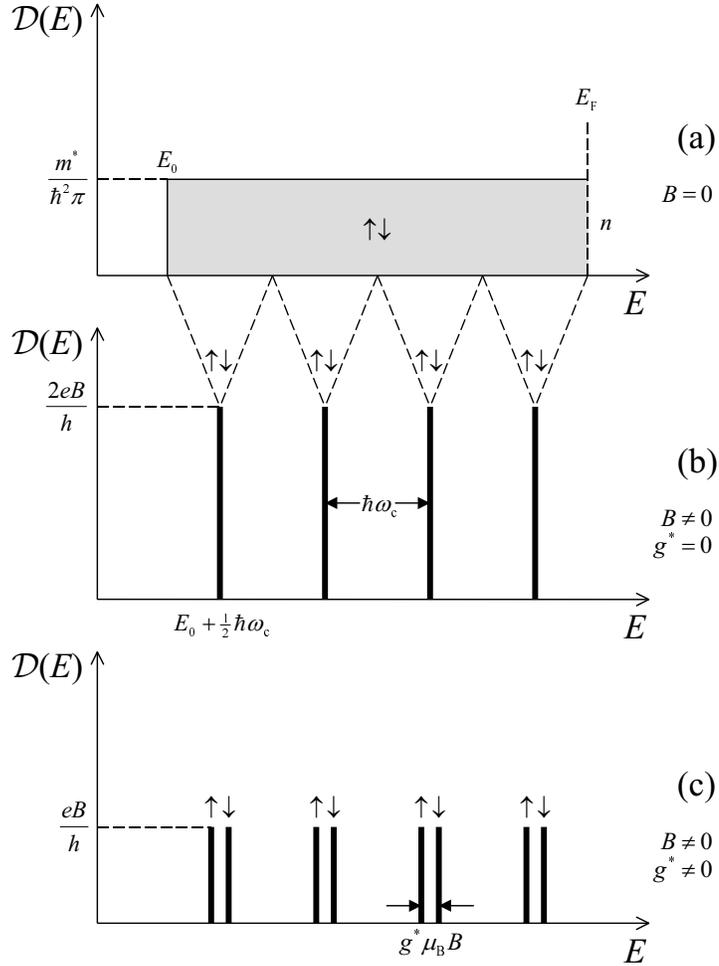}
\caption{Density-of-states as a function of energy for a 2D
carrier system: (a) in the absence of a magnetic field, (b) with a
magnetic field $(B)$ applied perpendicular to the 2D plane but
neglecting the spin-splitting of the resulting Landau levels, and
(c) with spin-splitting included. As is typical for a 2D electron
system in a standard, single GaAs/AlGaAs heterojunction, here it
is assumed that only one (size-quantized) electric subband, whose
edge energy is marked by $E_{o}$, is occupied.}
\end{figure}

The degeneracy of each spin-split quantized energy level is $eB/h$.
Since this degeneracy increases with $B$, to keep the total 2D
density of the system constant, the Fermi energy ($E_{F}$) has to
move so that fewer and fewer LLs are occupied with increasing $B$.
The number of spin-split LLs
occupied at a given $B$ is defined as the filling factor and is given by $%
\nu =n/\left( eB/h\right) =nh/eB$. Equivalently, $\nu $ is the
number of electrons per flux quantum $\Phi _{o}=h/e$. As $B$ is
increased and $E_{F}$ passes through the oscillating $D(E)$, nearly
all properties of the system, such as electrical resistivity,
magnetic susceptibility, heat capacity, etc., oscillate. The
magnetoresistance oscillations are often called Shubnikov-de Haas
oscillations. The oscillations are periodic in $1/B$ with frequency
$nh/e$ or $nh/2e$, depending on whether or not the spin-splitting is
resolved. This means that from a measurement of the frequency of the
oscillations one can deduce the density.

The delta-function-like energy levels shown in Fig. 6 are for an
ideally pure 2DES. In the presence of disorder, the levels are
broadened with their width, $\Gamma $, being of the order of
$\hbar/\tau _{q}$ where $\tau _{q}$ is the quantum lifetime of the
carriers. The states in the LLs' tails are \emph{localized} and only
the centers of the LLs contain current-carrying \emph{extended}
states. Now suppose the filling factor is $i$, or nearly $i$, so
that $E_{F}$ lies in the localized states between the $i$ and
($i+1$) LL. If the disorder and temperature are sufficiently small
so that $\Gamma $ and the thermal energy ($k_{B}T$) are both smaller
than the
LL separation, then as $T\rightarrow 0$ the longitudinal conductivity $%
\left( \sigma _{xx}\right) $ vanishes and the transverse or Hall
conductivity $(\sigma _{xy})$ becomes quantized
at a value that is equal to $ie^{2}/h$. This is the integral QHE. That $%
\sigma _{xx}\rightarrow 0$ is simply a consequence of there being no
extended states in the bulk of the 2D system to carry current. There
are, however, $i$ current-carrying ``edge states'' near the edge of
the sample (see the top left sketch in Fig. 2) and this leads to
$\sigma _{xy}$ being quantized although demonstrating this
quantization is more subtle (see, e.g., Refs. 6-8). Note also that,
according to the simple relations which convert the elements of the
conductivity tensor to those of the resistivity tensor, $\rho
_{xx}=\sigma _{xx}/\left( \sigma _{xx}^{2}+\sigma _{xy}^{2}\right) $
and $\rho _{xy}=\sigma _{xy}/\left( \sigma _{xx}^{2}+\sigma
_{xy}^{2}\right) $. Therefore, $\sigma _{xx}=0$ and $\sigma
_{xy}=ie^{2}/h$ mean that $\rho _{xx}=0$ and $\rho _{xy}=h/ie^{2}$.
This explains the experimental result in Fig. 1 for the Hall bar
sample shown in the inset.

To summarize, the IQHE is a consequence of: (1) the quantization
of the 2D system's energy levels into a set of well-defined (but
broadened) LLs with separation greater than $k_{B}T$, and (2) the
presence of localized states in between these LLs. Note that no
electron-electron interaction is needed to bring about or to
explain the IQHE.

\subsection{Electron-electron interaction and the fractional quantum Hall effect (FQHE)}

Suppose $B$ is sufficiently raised so that $\nu <1$. At $T=0$ the
kinetic energy of the 2DES is quenched and the system enters a
regime where, in the absence of disorder, its ground state is
determined entirely by the electron-electron interaction. In the
infinite $B$ limit, the system approaches a classical 2D system
which is known to be an electron crystal (Wigner crystal)
\cite{Grimes1979}. At finite $B$, the electrons cannot be localized
to a length smaller than the cyclotron orbit radius of the lowest
LL, or the \emph{magnetic length} $l_{B}=(\hbar /eB) ^{1/2}=(\nu
/2\pi n)^{1/2}$, and the ground state is typically a gas or liquid.
However, when $l_{B}$ is much smaller than the average distance
between electrons, i.e. when $\nu \ll 1$, a crystalline state is
possible \cite{WCreviews}. We will return to this crystalline state
in Section 3.4.

A competing ground state of the 2D system at high $B$ is the FQHE
liquid \cite{Tsui1982,Laughlin1983}. Ironically, the work that led
to the discovery of the new and totally unexpected FQHE phenomenon
\cite{Tsui1982} was itself one of the early experimental searches
for the magnetic-field-induced Wigner crystal! In high-quality,
low-disorder 2D carrier systems, in fact, the dominant ground states
of the system are the FQHE states (Fig. 7).  The FQHE, observed
at the principal fillings $\nu =1/q$ and other rational fractional fillings $%
\nu =p/q$ ($q$=odd integer) is characterized by the vanishing of
$\rho _{xx}$ and the quantization of $\rho _{xy}$ at $\left(
q/p\right) \left( h/e^{2}\right) $ as $T\rightarrow 0$. The effect
is phenomenologically similar to the IQHE but its origin is very
different. The FQHE state is an intrinsically many-body,
incompressible quantum liquid, described by the Laughlin
wavefunction:
\begin{equation}
\Psi _{m}^{\nu}\sim\prod_{i,j}(z_{i}-z_{j})^{m}\times \exp
\left(-\sum _i \frac{\left| z_{i}\right| ^{2}}{4l_{B}}\right).
\end{equation}
\noindent Here $z_{i}$ and $z_{j}$ are the (complex) coordinates of
pairs of electrons in the plane (Fig. 2, top center panel), and
$m=1/\nu$ is an \emph{odd} integer so that the wavefunction is
antisymmetric when two electrons are interchanged (Pauli exclusion
principle). Note also that the Coulomb repulsion between electrons
is built into this wavefunction as it becomes small when two
electrons come close to each other. The wavefunction has strong
short-range correlation but it does not describe a crystalline phase
as it has no long-range order (see e.g., the ``snap shots'' shown in
Fig. 7.7 of Ref. 6).
\begin{figure}\centering
\includegraphics*[bb=0 55 612.0 735.0,scale=0.5,width=5in,height=5in]{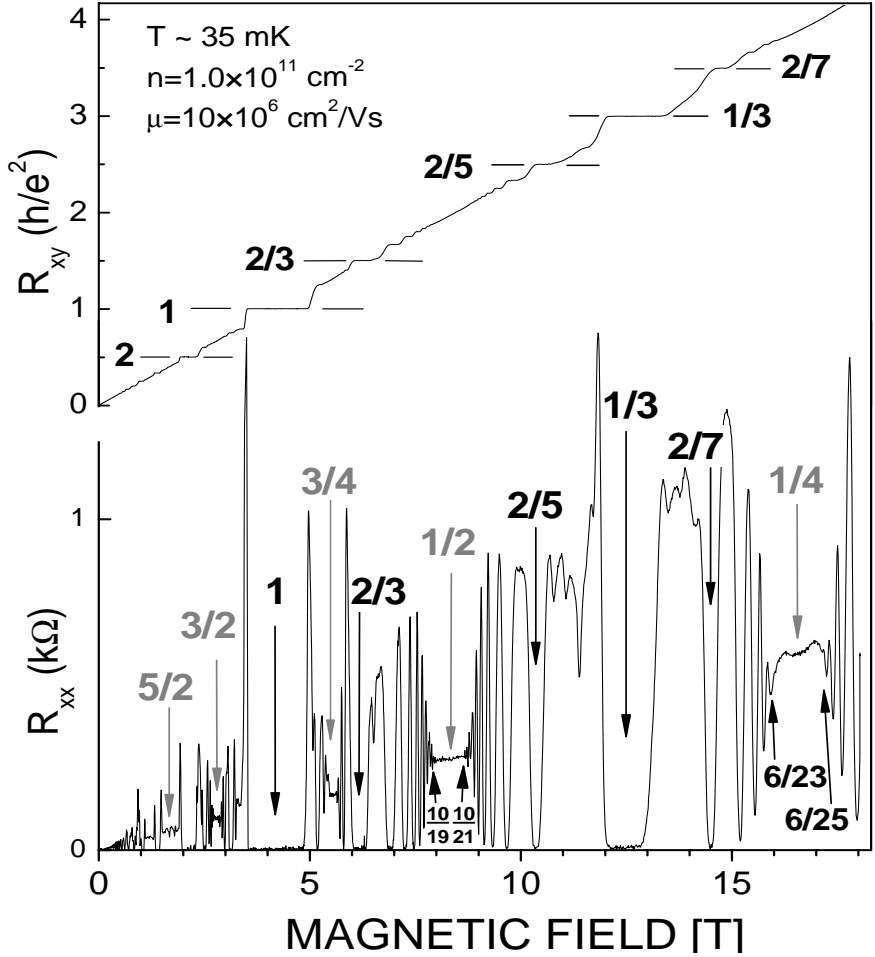}
\caption{Magnetoresistance traces for a very high quality
(mobility about $10^{7}$ cm$^{2}$/Vs) GaAs/AlGaAs sample, showing
numerous fractional quantum Hall states. (After Pan \emph{et al.}
\cite{Pan2002}.)}
\end{figure}

The FQHE has many fundamental and interesting characteristics among
which I briefly mention three here. First, the incompressibility
implies that the
ground state is separated from its excitations by a finite energy gap, $%
\Delta $. Experimentally $\Delta $ can be measured from the activated $T$%
-dependence of $\rho _{xx}$ according to $\rho _{xx}\sim \exp \left(
-\Delta /2k_{B}T\right) $. This is superficially similar to the
energy gap between the LLs which leads to the IQHE, but the origin
of the FQHE gap is entirely many-body. The theoretical $\Delta $ for
the $\nu =\frac{1}{3}$ FQHE in an \emph{ideal }2DES, with no
disorder, zero layer-thickness, and
infinitely separated LLs, is $\sim 0.1e^{2}/4\pi\varepsilon l_{B}$, where $%
\varepsilon $ is the dielectric constant of the host material
(GaAs). In \emph{real} samples, however, the finite-layer
thickness, LL mixing, and the ubiquitous disorder lead to a gap which is much smaller than $%
0.1e^{2}/4\pi\varepsilon l_{B}$ (see, e.g., Ref. 22). Finite layer
thickness, for example, leads to a softening of the short-range
component of the Coulomb interaction, and results in a weakening of
the FQHE. In fact, experiments \cite{Shayegan1990} and
calculations \cite{He1990} have revealed that once the layer thickness exceeds $%
\sim 3l_{B},$ the FQHE quickly collapses.

A second, quite intriguing yet fundamental feature of the FQHE is
that its elementary excitations carry fractional charge $e^{*}=e/m$.
There have been several reports of measuring this fractional charge,
one of the latest being measurements of the quantum shot noise which
is proportional to the charge of the conducting carriers
\cite{de-Picciotto1997}. In these measurements, the current noise
was monitored as a function of the backscattered current which
results from the tunneling between the FQHE edge states in a
point-contact (constriction). The results near $\nu =\frac{1}{3}$
indeed reveal that the current carrying particles have charge $e/3$.

The third noteworthy feature is the existence of FQHE states not
only at the primary fractional fillings such as $\nu =\frac{1}{3}$
and $\frac{1}{5}$, but also at a host of
other odd-denominator fillings. Examples are the states at $\nu =\frac{2}{5},%
\frac{3}{7},...$ and $\nu =\frac{3}{5},\frac{4}{7},....$ which can
be seen in Figs. 1 and 7. The strength of these states, namely their
measured energy gaps, typically decreases as the denominator of
their filling gets larger. Also, they appear to form a sequence of
decreasing strength as one goes from the primary state, such as $\nu
=\frac{1}{3}$, towards the even-denominator filling $\left( \nu
=\frac{1}{2}\right) $ at which there is no FQHE state. We will
revisit these observations in Section 3.3.  Note also that there are
unusual and not yet well-understood FQHE states at
\emph{even-denominator} fillings such as $\nu =\frac{5}{2}$ (see
Section 3.6).

\subsection{Composite Fermions}

The observation that the higher order FQHE states at fillings with
increasingly larger denominators have weaker strengths initially led
to an explanation for these states based on a ``hierarchical''
scheme where each state is considered the ``parent'' state for the
adjacent (in filling factor) weaker state. The idea is that as one
deviates from the exact filling for a given FQHE state,
quasiparticles are created above the energy gap; these
quasiparticles then interact and form an incompressible liquid once
their density to magnetic flux ratio reaches certain values. Such a
hierarchical construction can generate all the odd-denominator
fractions, and explains certain features of the observed FQHE
sequences. However, it fails to account for the observed
strength/weakness of all FQHE states. Also, in this scheme, the
wavefunctions of the higher order states turn out to be much more
complex than those for the primary states. Moreover, such
description of the FQHE differs entirely from that of the IQHE while
it is hard to overlook the striking similarity between the FQHE
sequence, e.g., at $\nu =\frac{1}{3},\frac{2}{5},\frac{3}{7}...$ and
the IQHE sequence at $\nu =1,2,3,...$ In fact, if we ``slide'' the
$5$.$5<B<9$ T portion of the
magnetoresistance trace in Fig. 1 to the left so that the position of $\nu =%
\frac{1}{2}$ is now the ``zero'' of the (effective) magnetic field,
we can see a one-to-one correspondence between the above IQHE and
FQHE sequences, both in terms of the field positions of $\rho _{xx}$
minima and their relative strength. (This is true if we assume that
the 2DES is fully spin-polarized at all fields.)

Such observations prompted the search for a description of the QHE
which somehow links the integral and fractional effects. This has
culminated in a remarkable description in which an even number of
fluxes combine with an electron to form a new, ``composite'' Fermion
(CF) \cite{CFreviews}. The electron-electron interaction and the
large magnetic field are embedded in this flux-electron
quasiparticle so that the system now behaves as if it contains
(essentially) non-interacting particles moving in an ``effective''
magnetic field which is the balance of the external field once the
attached fluxes are deducted. For example, focusing on the range
$\frac{1}{3}\leq \nu \leq \frac{1}{2}$ and attaching two fluxes to
each electron, the CF description maps the FQHE at $\nu
=\frac{1}{3},\frac{2}{5},\frac{3}{7},...$ to the IQHE at $\nu
=1,2,3,...$ Perhaps even more intriguing is the notion that the
effective field at $\nu =\frac{1}{2}$ is zero for the CFs so that,
at $\nu =\frac{1}{2}$, they ignore the large external magnetic field
and move about as if there is no magnetic field.

More rigorously, a gauge transformation that binds an even number of
magnetic flux quanta ($2l\Phi _{0}$ where $l$ is an integer and
$\Phi _{0}\equiv h/e$ is the flux quantum) to each electron maps the
2DES at even-denominator fillings to a system of CFs at a vanishing
$B_{eff}$ \cite{CFreviews}. Such transformation elegantly maps a
FQHE
observed at the 2DES filling $\nu $ to an IQHE for the CF system at filling $%
\nu ^{\prime }$ where $\nu ^{\prime }=\nu /\left( 1-2l\nu \right) $.
In addition, since $B_{eff}=0$ at $\nu =1/2l$, the CF system should
possess certain Fermi-liquid-like properties. Most notably, the CFs
should have a Fermi surface at $\nu =\frac{1}{2}$, just like
electrons do at zero magnetic field, and should therefore support
phenomena such as geometrical resonances and CF ballistic transport
\cite{CFreviews}.

Here I present, as an example, the results of an experiment which
demonstrate the surprisingly simple behavior of CFs near $\nu
=\frac{1}{2}$, namely their semiclassical, ballistic motion under
the influence of $B_{eff}$. Figure 8 shows data from a magnetic
focusing experiment \cite{Goldman1994} near $B=0$ (bottom trace) and
$\nu =\frac{1}{2}$ (top trace). The geometry of the experiment is
sketched in the inset, which shows the top view of the sample. Parts
of the sample are etched (thick lines in Fig. 8 inset) so that the
2DES is separated into three regions which are connected by two
narrow constrictions (point-contacts). The distance between the two
constrictions, \emph{L}, is chosen to be smaller than or of the
order of the mean-free-path of the electrons. Ballistic electrons
are then injected from the lower-left section to the upper section
through the ``injector'' constriction by passing a current between
the ohmic contacts marked +I and -I. Now a small $B$-field is
applied perpendicular to the plane to ``bend'' the semiclassical,
ballistic trajectory of the injected electrons as they travel in the
upper section. As $B$ is increased, whenever $L$ matches a multiple
integer of the electron's semiclassical cyclotron orbit diameter,
$d_c=2m^*v_F/eB=2\hbar k_F/eB$, the ballistic electrons impinge on
the ``collector'' constriction, either directly or after one or more
bounces off the focusing barrier separating the two constrictions.
At these $B$, one observes a maximum in the voltage measured between
the lower-right and the upper sections (contacts marked +V and -V).
The traces shown in Fig. 8 are the voltages measured between
contacts +V and -V, normalized to the current injected between
contacts +I and -I. Maxima can be clearly seen in the lower trace of
Fig. 8 for $B>0$ and their positions are indeed consistent with the
values of $L$ and $k_{F}$ for this 2DES. Note that for $B<0$, the
electrons are deflected to the left and no magnetic focusing is
expected, consistent with the absence of any observed maxima.
\begin{figure}\centering
\includegraphics[scale=0.5]{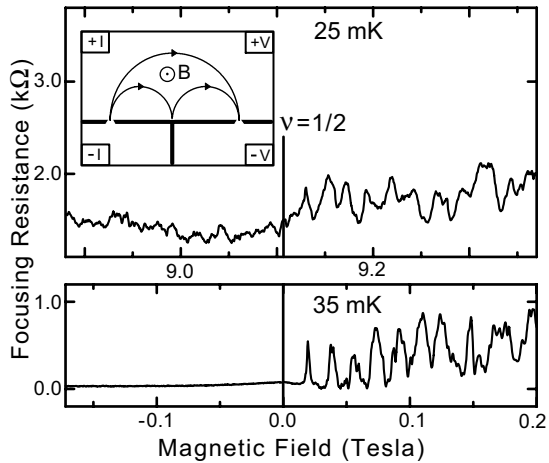}
\caption{Magnetic focusing spectra are shown for 2D electrons near
zero external magnetic field (bottom trace) and for composite
Fermions near $\nu =\frac{1}{2}$ (top trace) where the external
field is about 9.1 T. In the top trace, the position of $\nu
=\frac{1}{2}$ marks the zero of the effective magnetic field
$\left( B_{eff}\right) $ for CFs. Both traces exhibit peaks at
fields where the distance between the injector and collector
point-contacts ($L\simeq 5.3$ $\mu$m in this case) matches a
multiple integer of the classical cyclotron orbit diameter. The
inset schematically shows the top view of the sample. (After
Goldman \emph{et al.} \cite{Goldman1994}.)}
\end{figure}

The experiments of Goldman et al. \cite{Goldman1994} reveal
oscillations of the resistance not only near $B=0$ for electrons,
but also near $\nu =\frac{1}{2}$ (upper trace of Fig. 8). The data
provide a remarkable demonstration of the ``classical'', ballistic
motion of the CFs under the influence of $B_{eff}$. Note that
$B_{eff}$ is only a few tenths of a Tesla while the real external
magnetic field is about 12 Tesla! The large external magnetic flux
felt by the interacting electrons is replaced by the much smaller
flux influencing the apparently simple flux-electron composites. The
data of Fig. 8 also provide a direct determination of the Fermi
wavevector as well as an estimate for the ballistic mean-free-path
of the CFs $(\simeq 1$ $\mu$m).

As remarkable and perhaps non-intuitive as the CF picture may be, it
has received compelling verification through several key experiments
\cite{CFreviews}. These include measurements of the surface acoustic
wave propagation, FQHE activation energies, CF effective mass,
resistance oscillations in antidot arrays, magnetic focusing,
low-\emph{T} thermopower, magnetooptics, CF spin, temperature
dependence of the CF conductivity at $\nu =\frac{1}{2}$ and
$\frac{3}{2}$, and ballistic CF transport in nanostructures. The
results of most of these experiments are in general agreement with
each other and with the CF picture although some inconsistencies
exist. Among the unresolved topics are the CF effective mass and the
degree of CF spin polarization.

\subsection{The Wigner crystal state}
\begin{figure}[t]\centering
\includegraphics[scale=0.4]{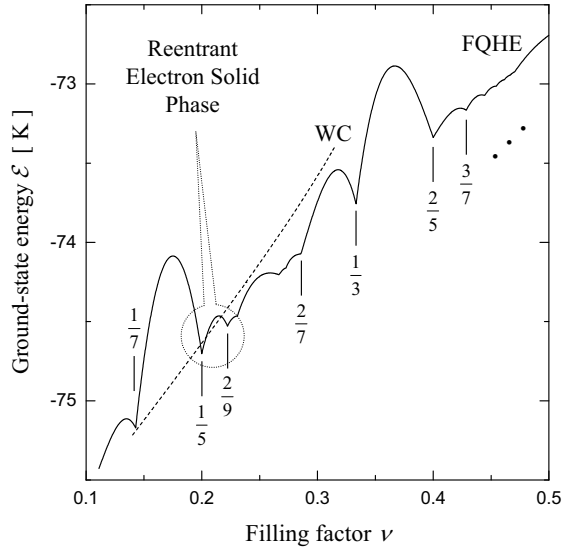}
\caption{Energies of two competing ground states of a 2D electron
system at high perpendicular magnetic fields. The FQHE
incompressible liquid states occur at special odd-denominator
fillings as the downward energy ``cusps'' indicate (solid curve).
The Wigner crystal (WC) state has monotically decreasing energy as a
function of inverse filling (dashed curve) and is expected to win
for fillings less than about $\frac{1}{6}$. (After Manoharan
\emph{et al.} \cite{Manoharan1997}.)}
\end{figure}
As mentioned earlier, the ground state of a 2DES at very high
magnetic fields, i.e., in the limit of $\nu \rightarrow 0$ is
expected to be a Wigner crystal (WC), namely, an ordered array of
electrons (see Fig. 2, top right panel).  A few words regarding the
competition between the FQHE and the WC are therefore in order. It
turns out that the Laughlin FQHE liquid states at $\nu =p/q$ are
particularly robust and have ground state energies which are lower
than the WC state energy, at least for $\nu
>\frac{1}{5}$ \footnote{This is not necessarily true for the higher Landau levels (see Section 3.6)}.
This is illustrated in Fig. 9 where the estimated energies are
plotted as a function of $\nu $ (for details of estimations see Ref.
28). The downward ''cusps" in energy reflect the incompressibility
of the FQHE states and the presence of energy gaps which are
proportional to the discontinuties in the derivative of energy vs
$\nu $. Also shown schematically in Fig. 9 (dashed curve) is the
expected dependence of the WC ground state energy on $\nu $.
Theoretical calculations predict that, in an ideal 2D system, the WC
should be the ground state for $\nu $ smaller than
about $\frac{1}{6}$. It is evident from Fig. 9 that while at $\nu =\frac{1}{5%
}$ the FQHE can be the ground state, the WC state may win as the
filling deviates slightly from $\frac{1}{5}$. It is possible
therefore to have a WC which is reentrant around a FQHE liquid
state.

The above picture has been used to rationalize the general current
belief that the insulating phase (IP) observed around the $\nu
=\frac{1}{5}$ FQHE in very high quality GaAs/AlGaAs 2DESs (e.g., see
Fig. 1) is the signature of a pinned WC state. The solid is
presumably ``pinned'' by the disorder potential, and can be made to
slide if a sufficiently large electric field is applied. Such
depinning would result in a nonlinear current-voltage characteristic
and various resonances, consistent with numerous measurements. The
magnetic-field-induced WC problem in 2D systems has been studied
extensively since the late 1980's \cite{WCreviews,Pan2002}.

High frequency measurements of the conductivity have proven to be a
valuable tool in the regime of small filling factors where the WC
phase is presumably dominant \cite{Ye2002}. An example is shown in
Fig. 10 for a very high quality GaAs 2DES. The data show a rather
sharp conductivity resonance at a frequency whose position and
characteristics, e.g., its behavior with temperature, density, and
magnetic field, are consistent with the pinning mode of a
magnetic-field-induced WC. Moreover, analysis of the resonance data
based on pinned WC models yields domain sizes that are many times
the inter electron spacing \cite{Ye2002}.
\begin{figure}\centering
\includegraphics[scale=0.5]{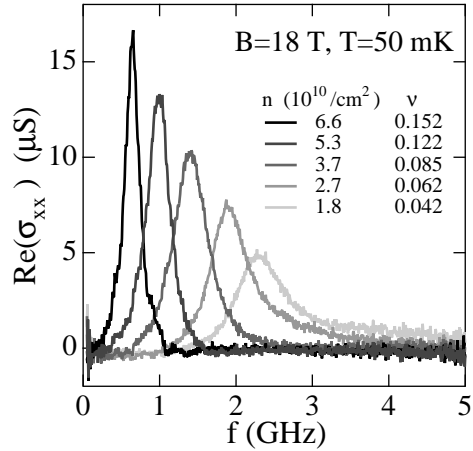}
\caption{Resonance spectra, showing the real part of the diagonal
conductivity as a function of frequency for a high mobility GaAs
2DES . Data are shown for a number of 2D densities at a fixed
field of 18 T. (After Ye \emph{et al.} \cite{Ye2002}.)}
\end{figure}

It is worth mentioning that strikingly similar reentrant IPs have
been observed in other high-quality GaAs 2D carrier systems, such as
in 2D \emph{hole} systems or in \emph{bilayer} systems of either
electrons or holes. As summarized in Fig. 11, however, the IPs in
these systems occur at much larger fillings. In a dilute 2D hole
system, e.g., an IP reentrant around the $\nu =\frac{1}{3}$ is
observed \cite{WCreviews}, while interacting {\it bilayer} electron
or hole systems with appropriate parameters show such phases at even
higher fillings (see Section 4.4). These observations can be
qualitatively understood in terms of the profound effect of Landau
level mixing (effective diluteness) in the case of the 2D holes and
of the interlayer interaction in the case of the bilayer systems,
both of which significantly modify the ground-state energies of the
FQHE and WC states of the system and shift the liquid-to-solid
transition to larger $\nu$ \cite{WCreviews}. These results are very
suggestive and provide further credibility to the interpretation of
the IP as a pinned WC.
\begin{figure}\centering
\includegraphics[scale=0.6]{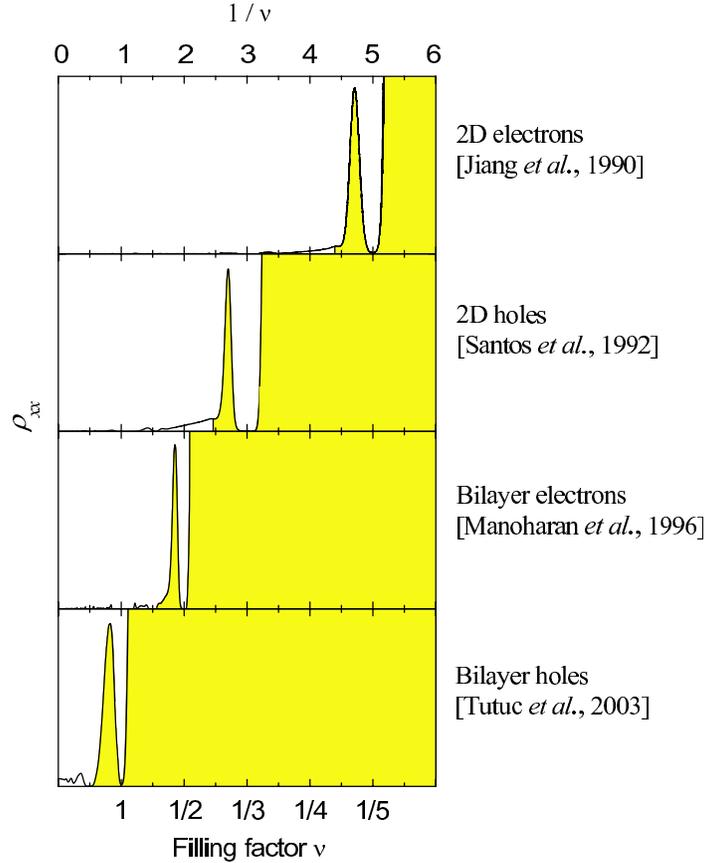}
\caption{Summary of reentrant insulating phases (shaded areas)
observed at low temperatures in various GaAs 2D carrier systems.
(After Jiang \emph{et al.} \cite{Jiang1990}; Santos \emph{et al.}
\cite{Santos1992}; Manoharan \emph{et al.} \cite{Manoharan1996};
Tutuc \emph{et al.} \cite{Tutuc2003}.)}
\end{figure}

\subsection{Ferromagnetic state at \textbf{$\nu = 1$} and Skyrmions}

For 2DESs in GaAs, while the IQHE at even $\nu $ arises from the
single-particle energy gaps separating the LLs, the spin splitting
of these levels leads to IQHE at odd $\nu $. The electron-electron
interaction and in particular the exchange energy, however, play a
dominant role for odd-$\nu $ IQHE and often lead to a substantially
larger QHE energy gap than expected from the bare effective
$g$-factor $\left( g^{*}\simeq -0.44\right) $ for GaAs
\cite{Nicholas1988}. In fact, according to theory \cite{Sondhi1993},
the odd-$\nu $ IQHE states should exist even in the limit of zero
Zeeman energy ($g^{*}\rightarrow 0$); there should be a spontaneous
ferromagnetic order with a spin polarized 2DES ground state.

Perhaps even more interesting are the predicted excitations of these
ferromagnetic states: provided that $g^{*}$ is sufficiently small,
the charged excitations of the system are finite-size ``Skyrmions",
termed so after the work of Skyrme in 1958 on baryons in nuclear
matter \cite{Skyrme1958}, rather than single spin flips. Skyrmions
are spin textures, smooth distortions of the spin field involving
several spin flips (see Fig. 2, bottom center panel)
\cite{Sondhi1993,Girvin1999}. The spin and size of the Skyrmions are
determined by the competition between the Zeeman and the exchange
energies: a large ratio of the exchange energy over the Zeeman
energy would favor large-size Skyrmions over single spin flips as
the (exchange) energy gained by the near parallelism of the spins
would outweigh the (Zeeman) energy cost of the extra spin flips.
Skyrmions are relevant at $\nu =1$ (at finite $T$) and near $\nu =1$
where the 2DES is not fully spin polarized.

Clear experimental evidence for finite-size Skyrmions was provided
by the pioneering nuclear magnetic resonance measurements of Barrett
{\it et al.} \cite{Barrett1995}. On either side of $\nu =1$, they
observed a rapid drop of the Knight-shift of the $^{71}$Ga nuclei
which are in contact with or are near the 2DES. Associating this
Knight-shift with the spin polarization of the 2DES, they deduced
that the charged excitations of the $\nu =1$ QHE carry large
($\simeq $4) spins \cite{Barrett1995}. This work has been followed
by numerous experimental and theoretical studies, providing
additional credence to the Skyrmionic picture near $\nu =1$
\cite{Girvin1999}.

Implied by the Knight-shift data \cite{Girvin1999,Barrett1995} is a
strong coupling of the nuclear and 2DES spin systems near $\nu =1$
where Skyrmions are present.
Here I would like to discuss some 2DES heat capacity $(C)$ data near $%
\nu =1$ at very low $T$ \cite{Bayot1996} which dramatically manifest
the consequences of this Skyrmion-induced coupling. Moreover, a
remarkably sharp peak observed in $C$ vs $T$ is suggestive of a
phase transition in the electronic system, possibly signaling a
crystallization of the Skyrmions at very low $T$.

Bayot {\it et al.} \cite{Bayot1996} succeeded in measuring $C$ vs
$B$ and $T$ in a GaAs/AlGaAs multiple-quantum-well sample in the QHE
regime down to $T\simeq 25$ mK (Fig. 12). Their $C$ vs $B$ data is striking in that at high $%
B$ (near $\nu =1$) $C$ becomes many orders of magnitude larger than its low $%
B$ value. While the low $B$ data can be understood based on the 2DES
electronic heat capacity and its oscillating density of states at
the Fermi energy, the high $B$ data near $\nu =1$ are unexpected and
cannot be accounted for based on the thermodynamic
properties of the 2DES alone. Both the very large magnitude of $C$ and the $%
T^{-2}$ dependence of $C$ at high $T$ (dashed line in Fig. 12) hint
at the nuclear Schottky effect \cite{Bayot1996}. Utilizing this
clue, Bayot \emph{et al.} were able to semi-quantitatively explain
the magnitude and the dependence of $C$ on $B$ and $T$ (for $T>0.1$
K) based on a simple Schottky model for the nuclear spins of the Ga
and As atoms in the quantum wells. Implicit in this interpretation
of course is a coupling between the nuclear spins and the lattice;
this coupling is assumed to be provided by the Skyrmions.

\begin{figure}\centering
\includegraphics*[scale=0.8]{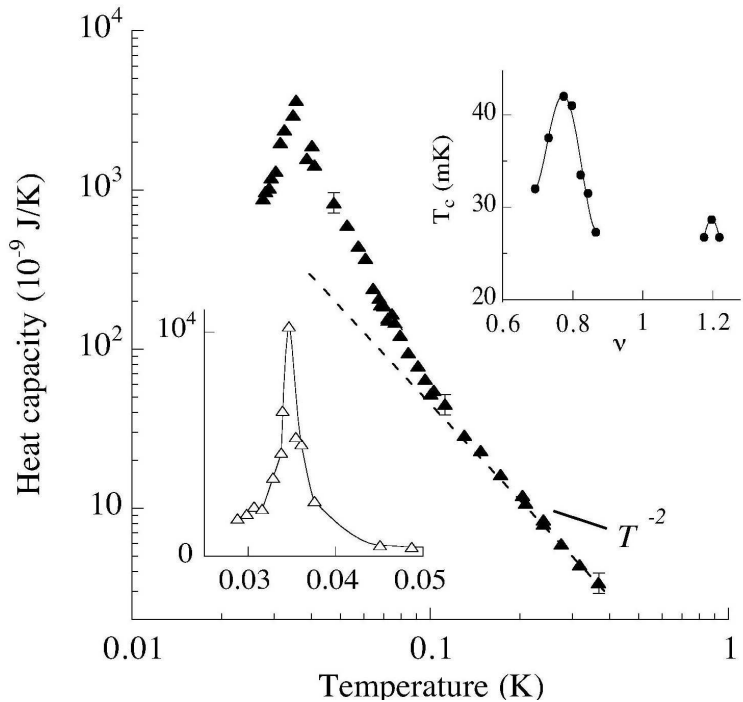}
\caption{Temperature dependence of the heat capacity near filling
factor one $(\nu =0.81)$ is shown in the main figure in a log-log
plot for a multiple-quantum-well GaAs sample. The dashed line shows
the $T^{-2}$ dependence expected for the Schottky model. The lower
inset shows a linear plot of the heat capacity vs temperature at
$\nu =0.85$. The temperature $T_{c},$ at which the heat capacity
exhibits the sharp peak depends on the filling factor as shown in
the upper inset. (After Bayot \emph{et al.} \cite{Bayot1996}.)}
\end{figure}

Figure 12 reveals yet another intriguing feature of the heat
capacity data: in a small range of $\nu $ near 0.8 (and also near
1.2), $C$ vs $T$ exhibits a very sharp peak at a temperature $T_{c}$
which sensitively depends on $\nu $ \cite{Bayot1996}. The Schottky
model, however, predicts a smooth maximum in $C$ at $T\sim \delta
/2k_{B}\simeq 2$ mK for $B\simeq 7$ T, the field position of
$\nu\simeq 0.8$, and cannot explain the sharp peak observed at
$T_{c}\sim 35$ mK ($\delta $ is the nuclear spin splitting). This
peak is possibly a signature of the expected Skyrmion
crystallization and the associated magnetic ordering near $\nu =1$
\cite{Bayot1996,Cote1997}. Such crystallization has indeed been
proposed theoretically \cite{Cote1997} although the details of the
Skyrmion liquid-solid transition and, in particular, how it would
affect the coupling to the nuclear spin system are unknown. One
feature of the data that qualitatively agrees with the Skyrmion
crystallization is worth emphasizing. As shown in the upper inset in
Fig. 12, the observed $T_{c}$ decreases rapidly as $\nu $ deviates
from 0.8 or 1.2 \cite{Bayot1996}; this is consistent with the
expectation that as the Skyrmion density decreases, the Skyrme
crystal melting temperature should decrease. Qualitatively
consistent with the heat capacity data and interpretation are the
results of a recent study of the nuclear magnetic resonance at very
low temperatures in a high-quality, single-layer, GaAs 2DES
\cite{Gervais2004}.

\subsection{Excited Landau levels: competition between uniform- and modulated-density many-body states}

The physics of 2D carrier systems in the second and higher LLs is
particularly delicate \cite{Eisenstein2001,Xia2004}. The electron
wavefunction in these excited levels has a larger extent and also
possesses one or more nodes. These combine to modify the
(exchange-correlation) interaction effects, and lead to a very close
competition between the uniform-density, liquid states and
density-modulated phases such as Wigner crystal or
charge-density-waves. Since these states are bunched together in
filling factor (e.g., $2 < \nu < 3$) and energy, their observation
requires the highest quality samples and lowest temperatures. Thanks
to the availability of ultra-high mobility GaAs 2DESs, it has become
possible in recent years to experimentally explore the high LLs in
detail and indeed many new phases have emerged.

Figure 13 highlights one of the novel features observed in high
quality 2D carrier systems at high LLs
\cite{Eisenstein2001,Lilly1999,Du1999,Shayegan2000}: the
longitudinal resistivity exhibits a very large in-plane anisotropy
at half-integer fillings $\nu \geq \frac{9}{2}$. The anisotropy
develops only at very low temperatures and can be as large as a
factor of ~10 in Hall bar geometry samples (the anisotropy is
exaggerated in van der Pauw geometry samples \cite{Simon1999}). Its
origin is believed to be the formation of interaction-induced
''stripe" phases at these half-filled LLs (see Ref. 42 for a brief
review). At $\nu =\frac{9}{2}$, e.g., the interaction leads to a
phase separation of the 2DES into stripes with $\nu = 4$ and 5
fillings (see Fig. 2, bottom right panel). Such striped,
charge-density-wave states were in fact theoretically predicted in
Hartree-Fock calculations
\cite{Kulakov1996,Fogler1996,Moessner1996}. The resistivity is small
(easy axis) when the current is passed parallel to the stripes and
large (hard axis) when the direction of the current is perpendicular
to the stripes. There is experimental evidence that the direction of
stripes can be rotated by adding a parallel component to the
magnetic field \cite{Pan1999,LillyCooper1999}. There is, however, no
clear picture yet as to what determines the direction of the stripes
in a purely perpendicular field
\cite{Eisenstein2001,Willett2001,Zhu2002}.
\begin{figure}\centering
\includegraphics[scale=0.6]{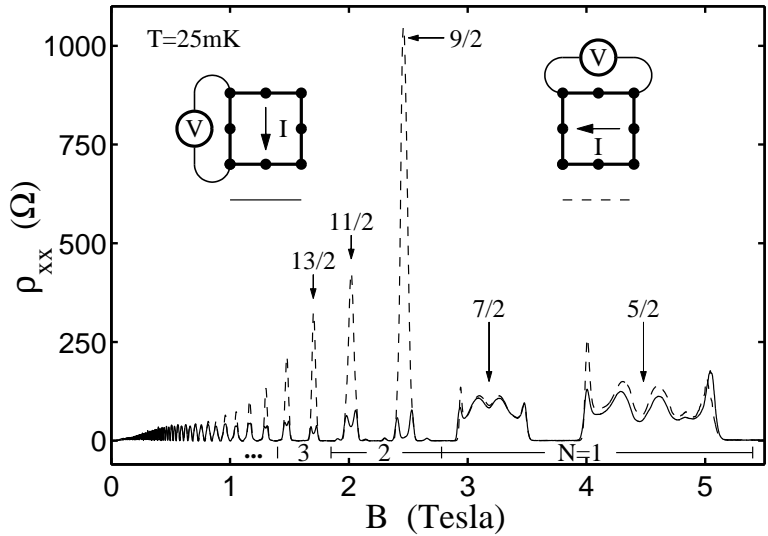}
\caption{Longitudinal magnetoresistance data for a high mobility
GaAs 2DES, revealing extreme anisotropy of $\rho _{xx}$ at
half-integer fillings $\nu \geq \frac{9}{2}$. The two traces
correspond to two different orientations of the current through the
sample (inest). (After Lilly \emph{et al.} \cite{Lilly1999}.)}
\end{figure}

Figure 14 provides yet another example of magnetotransport data
\cite{Xia2004} at $T=9$ mK of an extremely high-quality GaAs 2DES
with a mobility of $3.1\times10^7$ cm$^2$/Vs. Here data are shown in
the filling factor range $2 < \nu < 3$. Several FQHE states are
marked by arrows, including one at $\nu =\frac{5}{2}$. This
incompressible liquid state is very special since, unlike all the
other QHE states in a single-layer 2DES, it occurs at an {\it
even-denominator} filling. Recall that the odd-denominator rule is
linked to the requirement that the many-body FQHE wavefunction be
anti-symmetric (see equation 1). Although the $\frac{5}{2}$ FQHE
state was first observed quite some time ago \cite{Willett1987}, its
origin is not yet fully understood (for recent results see Refs.
56-58). In particular, it is still debated whether it is a
spin-polarized or unpolarized state \cite{PanStormer2001}, and also
whether a pairing of electrons or composite Fermions is responsible
for the formation of this state. The latest experimental data
suggest that there is a Fermi sea of composite Fermions at high
temperatures and that it is the pairing of these at the lowest
temperatures that leads to the $\frac{5}{2}$ FQHE
\cite{Willett2002}.
\begin{figure}\centering
\includegraphics[scale=0.2]{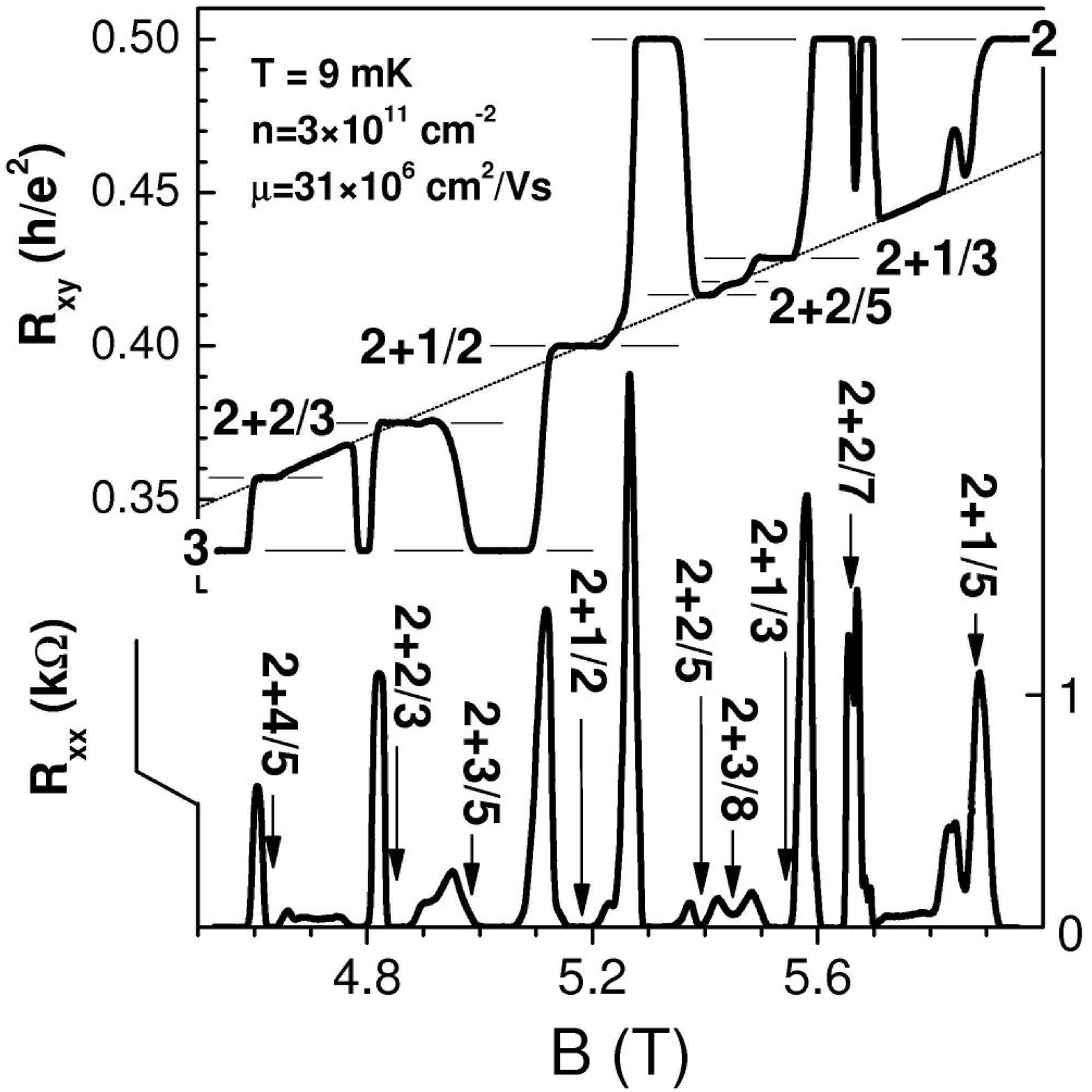}
\caption{Magnetotransport coefficients in the excited Landau level
(between $\nu=2$ and $\nu=3$) at a very low temperature of 9 mK for
a very high quality GaAs 2DES. The data reveal the complex and
delicate competition between FQHE states (marked by vertical arrows)
whose Hall resistance $R_{xy}$ is quantized at values on the
classical Hall line, and \emph{reentrant} IQHE states whose $R_{xy}$
is quantized at either two or three times $h/e^{2}$. (After Xia
\emph{et al.} \cite{Xia2004}.)}
\end{figure}

Another noteworthy feature of the data of Fig. 14 is that in certain
ranges of magnetic field, e.g., between 5.0 and 5.1 T, $R_{xx}$
vanishes while the Hall resistance $R_{xy}$ attains a quantized
value equal to the neighboring \textit{integer} QHE plateaus
($3h/e^2$ in this example) \cite{Xia2004,Eisenstein2002}. These so
called {\it reentrant} QHE states exhibit remarkable non-linear
current-voltage characteristics and narrow-band noise
\cite{Cooper2003}, as well as pronounced resonances at microwave
frequencies \cite{Lewis2002,Chen2003}. The data all signal that
these are non-uniform density, many-body states and, together with
theoretical results, suggest the presence of exotic pinned Wigner
crystal and ''bubble" phases (a Wigner crystal phase with more than
one electron in the unit cell).

\subsection{Radiation-induced ``zero-resistance'' states at low fields}

This is one of the latest developments in the physics of 2DESs in a
perpendicular magnetic field
\cite{Zudov2001,Ye2001,Mani2002,Zudov2003}. Figure 15 shows the
basic observation \cite{Mani2002}. When a very high quality 2DES is
irradiated with microwaves ({\it f} = 50 GHz for Fig. 15 data) , its
longitudinal resistance develops deep oscillations at very low
magnetic fields, corresponding to $\nu
> 50$. The resistance minima get stronger and become vanishingly
small as the temperature is lowered toward absolute zero.  Similar
to the Shubnikov de Haas effect, these oscillations are periodic in
$B^{-1}$, but the positions of the minima are not determined by the
chemical potential and the filling factor, but rather by the ratio
of the microwave and cyclotron frequencies. This observation implies
that these oscillations have a semi-classical origin. Moreover, as
seen in Fig. 15, there are no plateaus in the Hall resistance so the
phenomenon is distinct from the QHE.
\begin{figure}\centering
\includegraphics[scale=0.4]{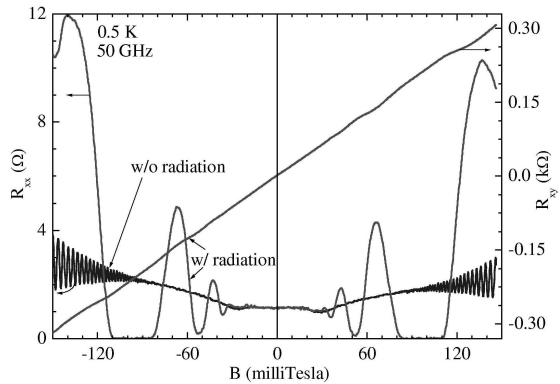}
\caption{Magnetotransport coefficients of a high mobility GaAs 2DES
under 50 GHz incident radiation. Under radiation, at low magnetic
fields, the longitudinal resistance $R_{xx}$ develops deep minima
that approach zero at low temperatures but no plateaus are formed in
the Hall resistance $R_{xy}$. (After Mani \emph{et al.}
\cite{Mani2002}.)}
\end{figure}

The observation of these so-called radiation-induced
''zero-resistance" states has generated an enormous number of
theoretical papers (see \cite{Durst2004} for a partial list of some
of the theory papers). While it is premature to say that the problem
is entirely understood, the plausible picture that appears to be
emerging is that this is not a new collective phenomenon, but rather
a manifestation of non-equilibrium dynamics under the influence of
microwave radiation. In particular, a simple model, based on the
effect of microwave radiation on impurity scattering can explain the
period and phase of the resistance oscillations
\cite{Durst2004,Durst2003,Ryzhii1970,Ryzhii1986}. The model in fact
predicts \emph{negative} resistance minima, i.e., the amplitude of
the oscillations should grow with the intensity of the microwaves so
that, at sufficiently large intensity, the resistance should become
negative in certain ranges of the magnetic field. Experimentally,
however, the resistance minima appear to saturate as they approach
zero (Fig. 15). A theoretical explanation for this saturation has
been proposed based on current instabilities associated with the
negative resistance \cite{Andreev2003,Vavilov2004}. The explanation
asserts that a uniform current is unstable if $R_{xx} \rightarrow
0$, and that the sample breaks into domains, in some of which the
current flows opposite to the applied current. Such domains are yet
to be detected in measurements.

\section{Correlated bilayer electron states}

\subsection{Overview}

The introduction of an additional degree of freedom can have a
profound effect on the many-body ground states of the 2DES at high
\emph{B} \cite{BilayerReview}. For example, the addition of a spin
degree of freedom stabilizes particular spin-unpolarized FQHE
observed at lower \emph{B} \cite{Clark1989}, while substantially
increasing the layer thickness (thus introducing an additional
spatial degree of freedom) leads to a weakening and eventual
collapse of the FQHE \cite{Shayegan1990,He1990}.  On the other hand,
adding a \emph{layer} degree of freedom can lead to novel correlated
states some of which have no counterpart in single-layer 2DESs.  For
example, when two electron layers are brought to close proximity so
that the interlayer and intralayer Coulomb interactions are
comparable, new QHE states ensue.  Such ''two-component" QHE states
have a generalized Laughlin wavefunction of the form
\cite{Halperin1983,Yoshioka1989,MacDonald1990}:

\begin{equation}\label{2}
\Psi^\nu_{mml}\sim\prod_{i,j}(u_i-u_j)^m\prod_{i,j}(w_i-w_j)^m\prod_{i,j}(u_i-w_j)^l\times
\exp \left(-\sum_i\frac{\left|u_i\right|^2}{4l_B^2}
-\sum_i\frac{\left|w_i\right|^2}{4l_B^2}\right)
\end{equation}

\noindent where $ u_{i}$ and $w_{i}$ denote the complex 2D
coordinates of the electrons in the two layers. The integer
exponents $m$ and $l$ determine the intralayer and interlayer
correlations, respectively, and the total filling factor for the
$\Psi _{mml}^{\nu }$ state is $\nu =2/\left( m+l\right)$.

A measure of how close one needs to bring two electron layers for
novel bilayer phenomena to occur is the ratio of the intralayer and
interlayer Coulomb interaction, $\left(e^{2}/4\pi\varepsilon
l_{B}\right) /\left(e^{2}/4\pi\varepsilon d\right) = d/l_{B}$, where
$d$ is the interlayer distance and $l_{B}=(\hbar/eB)^{1/2}$ is the
magnetic length. Now the two-component QHE states described by $\Psi
_{mml}^{\nu }$ come in two classes. For large $d/l_{B}$, the system
behaves as two independent layers in parallel, each with half the
total density. The FQHE states in this regime therefore have {\it
even numerator} and odd denominator. An example is the $\Psi
_{330}^{2/3}$ state which has a total filling of $\frac{2}{3}$
($\frac{1}{3}$ filling in each layer). Note that the exponent $l=0$
means that there is no interlayer correlation. For small enough
$d/l_{B}$, on the other hand, fundamentally new QHE states with
strong interlayer correlation are possible. Two such states that
have been observed so far are the $\Psi _{331}^{1/2}$ and $\Psi
_{111}^{1}$ states at $\nu =\frac{1}{2}$ and at $\nu =1$,
respectively.  I will briefly present these states in the next two
sections and then discuss another phenomenon, namely the presence of
insulating phases at high filling factors, that suggests the
formation of \emph{bilayer} Wigner crystal states.

Before presenting data, it is worth noting that a high quality,
GaAs/AlGaAs-based bilayer electron (or hole) system can be realized
in two distinct types of structures. One is a double quantum well
(QW) structure where the electrons are confined to two GaAs wells
that are separated by an AlGaAs barrier (Fig. 16, left side). A
second, less intuitive structure is a single, wide GaAs QW of width
$\sim$ 100 nm. At low density the electrons occupy the lowest
electric subband and have a single-layer-like (but rather ``thick''
in the growth direction) charge distribution. As more electrons are
added to the well, their electrostatic repulsion forces them to pile
up near the well's sides and the resulting electron charge
distribution appears increasingly bilayer-like (Fig. 16, right
side). The wide QW system is particularly interesting since both the
inter-layer tunneling and, to some degree, the distance between the
layers, can be tuned {\it in situ} by adding or removing electrons
from the QW (via the application of front- and back-sides)
\cite{Manoharan1996,Suen1991,Shayegan1999}. This means that in a
wide QW of fixed width, the system can be essentially tuned from a
bilayer to a (thick) single-layer by decreasing the density. This
evolution with density plays a decisive role in the properties of
the correlated electron states in this system
\cite{Manoharan1996,Suen1991,Shayegan1999}.
\begin{figure}\centering
\includegraphics[scale=0.7]{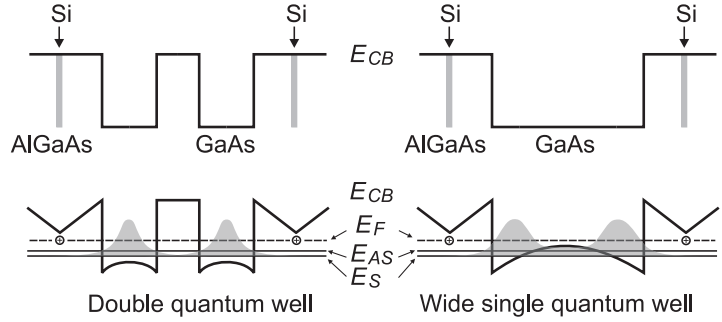}
\caption{Schematic figure showing the formation of a bilayer
electron system in either a double quantum well (left) or a wide
quantum well (right). In each case, the conduction band edge is
shown before (top) and after (bottom) the charge transfer from the
dopants to the quantum well(s). Note that in the wide quantum well
case, the "barrier" between the layers results from the
(self-consistent) electrostatic repulsion between the electrons
themselves.}
\end{figure}

\subsection{Even-denominator quantum Hall states in bilayer systems}

Figure 17 exhibits data for an electron system in a 75 nm-wide,
single GaAs quantum well. As seen in the main panel, there is a
well-developed FQHE state at the \emph{even-denominator} filling
$\nu =\frac{1}{2}$ \cite{Suen1992}. A similar FQHE state has also
been observed in GaAs double QW samples \cite{Eisenstein1992}. The
inset to Fig. 17 reveals a FQHE state at yet another
even-denominator filling $\nu =\frac{3}{2}$ \cite{Suen1994}. Neither
of these even-denominator states has a counterpart in standard 2DESs
in single-heterostructures. I emphasize that the fillings marked in
Fig. 17 are the \emph{total} fillings of the bilayer system; e.g.,
$\nu =\frac{1}{2}$ corresponds to $\frac{1}{4}$ filling for each
layer.
\begin{figure}\centering
\includegraphics[scale=0.5]{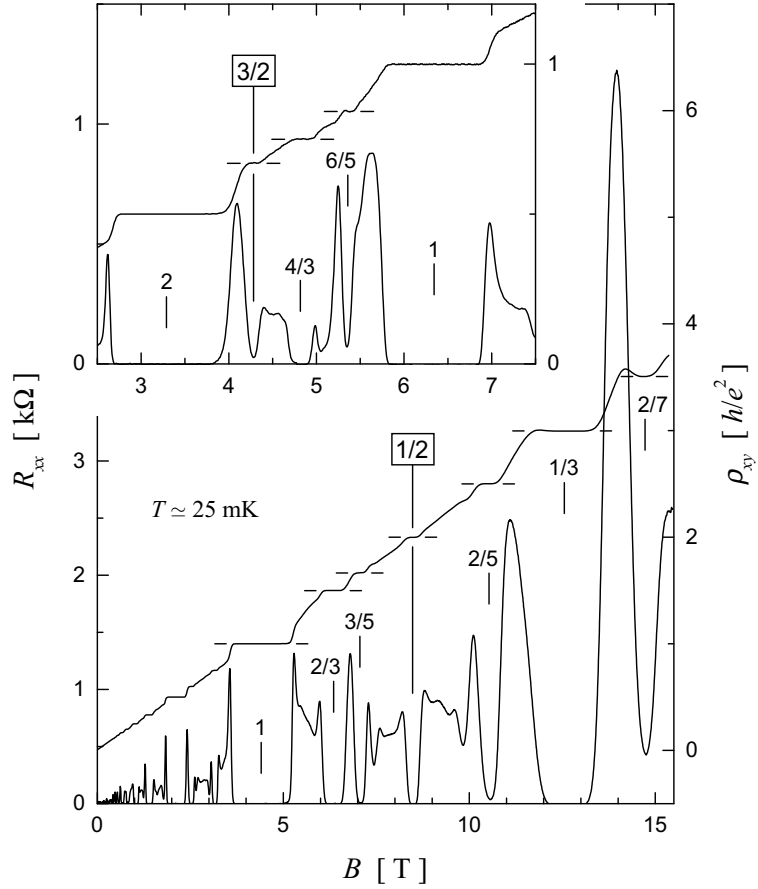}
\caption{Magnetotransport data, taken at $T\simeq 30$ mK, for a 75
nm-wide, single GaAs quantum well with $n=1.03\times 10^{11}$
cm$^{-2}$ (main figure) and $n=1.55\times 10^{11}$ cm$^{-2}$
(inset), showing well-developed \emph{even-denominator} FQHE states
at $\nu =\frac{1}{2}$ and $\frac{3}{2}$, respectively. These unique
FQHE states are stabilized by both interlayer \emph{and} intralayer
correlations. (After Suen \emph{et al.} \cite{Suen1992}.)}
\end{figure}

The FQHE states at $\nu =\frac{1}{2}$ observed in the double QW and
the wide single QW structures are believed to be signatures of the
$\Psi _{331}^{1/2}$ state. (The state at $\nu =\frac{3}{2}$ can be
understood as the hole conjugate of the $\frac{1}{2}$ state.) In the
double QW system, the FQHE state at $\nu =\frac{1}{2}$ is observed
for $d/l_B\simeq 2$, consistent with theoretical expectations for
the $\Psi _{331}$ state. In the wide QW system, on the other hand,
the $\frac{1}{2}$ state is stable at much larger $d/l_B$ values
($\sim$ 6). This is likely because the larger thickness of the
electron layers in the wide QW leads to a softening of the
intralayer interaction in this system \cite{Suen1994}.

\subsection{The bilayer QHE at \textbf{$\nu = 1$}: electron-hole pairing and Bose-Einstein condensation}
\begin{figure}\centering
\includegraphics[scale=0.5]{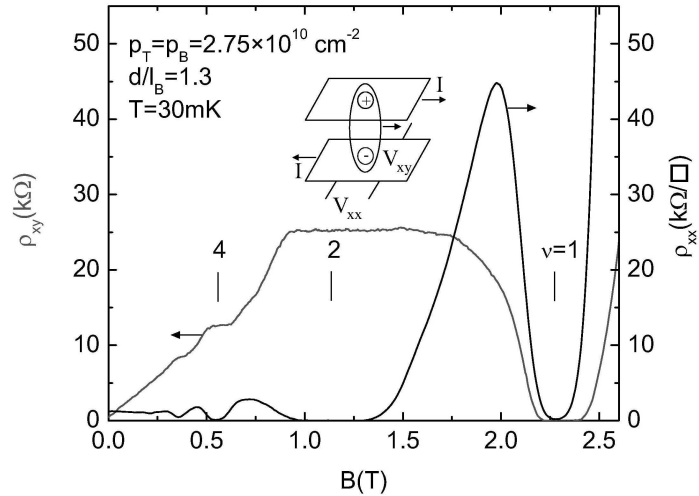}
\caption{Magnetotransport coefficients of a bilayer 2D hole system
in the \emph{counterflow} geometry (inset) where the current is
passed through the two layers in opposite directions. Both the
longitudinal \emph{and} the Hall resistivity vanish in the $\nu =1$
QHE state. The vanishing of the Hall resistivity signals the
formation of bilayer electron-hole pairs (excitons), shown
schematically in the inset. (After Tutuc \emph{et al.}
\cite{Tutuc2004}.)}
\end{figure}
In closely spaced bilayer systems, the interlayer and intralayer
interactions can also lead to the $\Psi _{111}$ QHE state at total
filling factor $\nu = 1$
\cite{Chakraborty1989,Wen1992,Murphy1994,Moon1995}. As can be seen
from equation (2), the exponents of the three terms in the $\Psi
_{111}$ wavefunction are all equal, meaning that this state enjoys
similar interlayer and intralayer correlations. As a result, this is
a very special state: it possesses unique, interlayer phase
coherence that leads to exotic properties such as electron-hole
pairing and Bose-Einstein condensation \cite{Wen1992,Moon1995}. To
understand the physics of this peculiar QHE state note that, at $\nu
= 1$, the carriers in each layer occupy exactly half of the
available Landau orbits, leaving the other half vacant. Under proper
circumstances, namely at low temperatures and when $d/l_B \simeq 1$,
the carriers in one layer ''pair" with the vacancies in the opposite
layer and form neutral excitons. These excitons in turn can condense
into a superfluid state below a critical temperature. Alternatively,
one can ascribe the layer degree of freedom in this system to a
pseudo-spin. The system's ground state is then a QHE ferromagnet
where all the pseudo-spins align in the same direction.

Experimental data have already shown that the bilayer $\nu =1$ QHE
exhibits novel phenomena such as Josephson-like interlayer tunneling
\cite{Spielman2000} and quantized Hall drag \cite{Kellogg2002}. Here
I present an example of the fascinating phenomenon recently observed
in this system, namely electron-hole pairing and signatures of
superfluidity in the ''counterflow" transport configuration
\cite{Kellogg2004,Tutuc2004,Wiersma2004}. Figure 18 shows such
counterflow data for a bilayer GaAs \emph{hole} system
\cite{Tutuc2004}; qualitatively similar data have also been reported
for bilayer electron systems \cite{Kellogg2004,Wiersma2004}. In the
counterflow geometry, equal currents are passed in opposite
directions in the two, independently contacted layers (see inset to
Fig. 18). At $\nu = 1$, both the longitudinal \emph{and} the Hall
counterflow resistances tend to vanish in the limit of zero
temperature. The vanishing of the Hall resistivity is especially
important since it directly demonstrates that the counterflow
current is carried by \emph{neutral} particles, that is, by
particle-vacancy pairs which have zero electric charge and therefore
experience no Lorentz force. The vanishing of the longitudinal
resistivity in the limit of zero temperature implies that the ground
state of the system is an excitonic Bose-Einstein condensate
(superfluid). It is worth emphasizing that the longitudinal
\emph{conductivity} $\sigma_{xx}$, determined by inverting the
measured longitudinal and Hall resistivities, \emph{diverges} as
{\it T} $\rightarrow 0$. This is consistent with a superfluid ground
state, and opposite to what is observed in the usual QHE where the
finite $\rho_{xy}$ leads to a vanishing $\sigma_{xx}$ in the limit
of zero temperature. So far, however, a transition to the superfluid
state at finite temperature has not been observed experimentally,
possibly because of residual sample disorder.

\subsection{Insulating phases in bilayer systems: evidence for a bilayer Wigner crystal}
\begin{figure}[h]\centering
\includegraphics[scale=0.5,width=4.5in,height=4.5in]{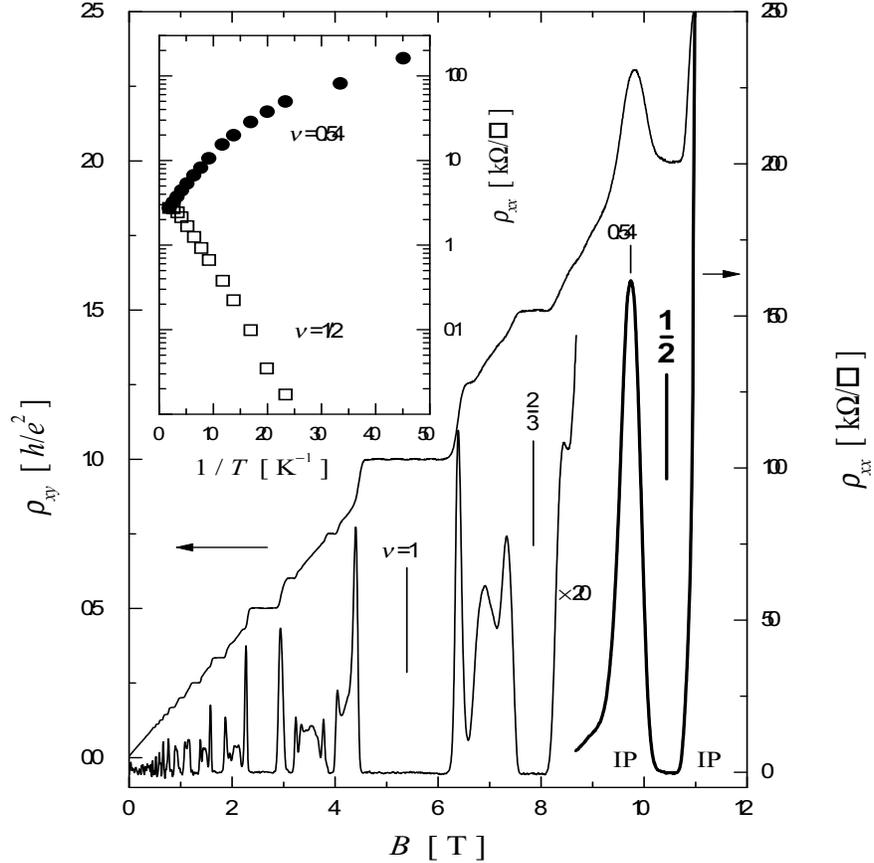}
\caption{Data for the same sample of Fig. 17 but with $%
n=1.26\times 10^{11}$ cm$^{-2}$. Here we an insulating phase
reentrant around the $\nu =\frac{1}{2}$ FQHE state is observed. The
inset shows
the temperature dependence of resistivity: at $\nu =\frac{1}{2}$, $\rho _{xx}$ vanishes as $%
T\rightarrow 0$ indicative of a FQHE state while at slightly higher
and lower $\nu $, $\rho _{xx}$ shows an insulating behavior as it
diverges with decreasing $T$. (After Manoharan \emph{et al.}
\cite{Manoharan1996}.)}
\end{figure}
Figure 19 reveals yet another interesting observation in interacting
bilayer systems, namely, the development of insulating phases (IPs)
that are reentrant around FQHE states at rather high fillings. In
Fig. 19, transport data on the same wide QW as in Fig. 17 but at a
higher density ($n\simeq 1.26\times 10^{11}$ cm$^{-2}$) show IPs
reentrant around the $\nu = \frac{1}{2}$ FQHE state. Note that $\nu
=\frac{1}{2}$ means a layer filling of $\frac{1}{4}$, i.e., a
filling which is larger than $\frac{1}{5}$ where an IP is observed
in single-layer, GaAs 2DESs (see Fig. 11 in Section 3.4).
Presumably, the interlayer interaction is leading to a
\emph{bilayer}, pinned Wigner crystal. The evolution of this
insulating phase as a function of total bilayer density and layer
density imbalance was found to be consistent with this conjecture
\cite{WCreviews,Manoharan1996,Shayegan1999}.

Experimental data on interacting, GaAs, bilayer \emph{hole} systems
further corroborate this conclusion. In the case of holes, an IP
reentrant around $\nu =1$ is observed (Fig. 11)
\cite{Tutuc2003,Faniel2005}. Here the IP is seen around the bilayer
($\Psi _{111}$) QHE state, i.e., near a layer filling of
$\frac{1}{2}$ which is larger than $\nu =\frac{1}{3}$ where the IP
in a single-layer 2D hole system occurs (see Fig. 11), again
suggesting that interlayer interaction has shifted the onset of the
Wigner crystal formation to higher fillings.

\section{Summary and future perspectives}

In this article I have attempted to provide a glimpse of the
exciting phenomena that 2D carrier systems in a perpendicular
magnetic field have revealed over the last 25 years or so. With
improvements in sample quality, it is more than likely that new
surprises continue to emerge. This is particularly true for the
excited Landau levels (Section 3.6) where there is fierce
competition between various uniform- and modulated-density many body
states. Higher quality samples and lower temperatures are key to the
observation and further understanding of such states.

A second area where more surprises are likely to emerge is in
studies of 2D carrier systems in novel structures and materials.
Examples are the bilayer or, more generally, \emph{multicomponent}
carrier systems. Such systems possess an additional degree of
freedom, e.g., \emph{layer}, \emph{spin}, or \emph{valley} degree of
freedom, and this can lead to phenomena that at times have no
counterpart in one-component systems (see Section 4 for examples).
Recently there has also been progress in the fabrication of
high-quality multi-valley systems, including 2D electrons confined
to AlAs \cite{DePoortere2002} or Si \cite{Lai2004} quantum wells,
and the new systems have indeed revealed intriguing FQHE phenomena
stemming from their multiple valley occupation. An example is shown
in Fig. 20 for an AlAs quantum well where the 2D electrons occupy
two conduction band valleys \cite{DePoortere2002}. A developing FQHE
state is observed at a very high filling of $\nu =\frac{11}{3}$, and
there are also hints of FQHE states emerging at higher fillings
(e.g., $\nu =\frac{13}{3}$, $\frac{14}{3}$, and even $\frac{17}{3}$
when the sample is tilted with respect to the direction of magnetic
field \cite{DePoortere2002}). Fractional QHE states at such high
fillings are either absent or rarely seen in standard, GaAs 2DESs of
even the highest quality, and are likely a result of the
multi-valley electron occupation.
\begin{figure}\centering
\includegraphics[scale=0.8]{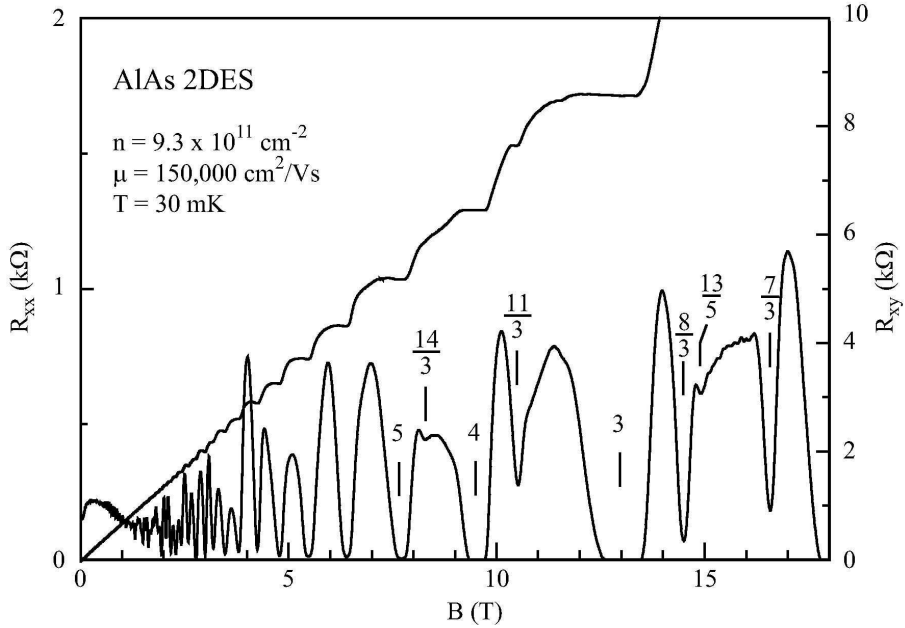}
\caption{Magnetotransport data for 2D electrons occupying two
conduction band valleys in an AlAs quantum well. The data exhibit
developing FQHE states at unusually large fractional fillings such
as $\nu =\frac{11}{3}$. (After De Poortere \emph{et al.}
\cite{DePoortere2002}.)}
\end{figure}

In closing, I would like to emphasize that the focus of this paper
has been transport measurements on 2D carrier systems in a
\emph{perpendicular} magnetic field. Much can be learned about the
physics of 2D systems by either adding an in-plane component of the
magnetic field, or applying a purely in-plane field. For example, at
appropriate tilt angles, one can bring the Landau levels of opposite
spin into coincidence and study phenomena such as \emph{quantum Hall
ferromagnetism} \cite{Jungwirth2000,DePoortere2000}. Or by applying
a purely in-plane field, one can study the spin polarization of 2D
systems and determine their spin susceptibility (see, e.g., Ref. 97
and references therein).

Finally, magnetic fields are also invaluable tools in studying
systems with dimensions lower than two.  Examples include quantum
\emph{wire} (1D) and \emph{dot} (0D) systems. In such systems, the
magnetic field can couple to the spin of the carriers and, for
appropriate geometries and parameters, also to their orbital motion.
These systems are the subjects of intensive current research, thanks
to their exciting basic properties as well as potential device
applications \cite{EP2DSconfs}.

\section{Acknowledgments}

I thank all the colleagues in the field of 2D carrier systems in a
magnetic field for years of hard work, illuminating discussions,
fruitful collaborations, and exciting competition! I apologize for
not giving adequate credit to everyone's work and for being biased
toward work done in my lab with which I am most familiar. Many
thanks to L.W. Engel, S.M. Girvin, A.H. MacDonald, D.C. Tsui, and E.
Tutuc, for a critical reading of this article. I am indebted to E.P.
De Poortere, L.W. Engel, V.J. Goldman, O. Gunawan, M.P. Lilly, R.
Mani, H.C. Manoharan, S. Melinte, W. Pan, Y.P. Shkolnikov, and E.
Tutuc for providing me with figures, including some unpublished
ones. I also thank Y.P. Shkolnikov, O. Gunawan, and S. Braude for
help preparing the manuscript. The work at Princeton University has
been supported primarily by the National Science Foundation and the
Department of Energy.

\end{document}